\documentclass[twocolumn]{aastex631}

\usepackage{CJK}
\usepackage{mathtools}

\usepackage{enumitem}
\usepackage{url}

\usepackage[most,breakable]{tcolorbox}

\usepackage{xcolor}

\definecolor{gold}{RGB}{255,215,0}
\definecolor{silver}{RGB}{192,192,192}
\definecolor{bronze}{RGB}{205,127,50}

\definecolor{boxcolor1}{RGB}{230,240,255} 
\definecolor{boxcolor2}{RGB}{255,240,230} 
\definecolor{boxcolor3}{RGB}{230,255,230} 
\definecolor{boxcolor4}{RGB}{255,230,255} 
\definecolor{boxcolor5}{RGB}{255,255,230} 
\definecolor{boxcolor6}{RGB}{230,255,255} 
\definecolor{grey}{RGB}{240,240,240} 

\newcommand{\goldmedal}{\textcolor{gold}{\LARGE\textbullet}}
\newcommand{\silvermedal}{\textcolor{silver}{\LARGE\textbullet}}
\newcommand{\bronzemedal}{\textcolor{bronze}{\LARGE\textbullet}}
\newcommand{\notable}{\textcolor{cyan!60}{\LARGE\textbullet}}

\newtcolorbox{questionbox}[1][]{
  colback=#1,
  colframe=black,
  boxrule=1pt,
  arc=5pt,
  outer arc=5pt,
  enhanced,
  breakable
}

\shorttitle{Astronomy Jeopardy!}
\shortauthors{Ting et al.}

\begin{document}
\begin{CJK*}{UTF8}{gbsn}

\title{AstroMLab 1: Who Wins Astronomy Jeopardy!?}

\author{Yuan-Sen Ting (丁源森)}
\affiliation{Research School of Astronomy \& Astrophysics, Australian National University, Cotter Rd., Weston, ACT 2611, Australia}
\affiliation{School of Computing, Australian National University, Acton, ACT 2601, Australia}
\affiliation{Department of Astronomy, The Ohio State University, Columbus, OH 43210, USA}
\affiliation{Center for Cosmology and AstroParticle Physics (CCAPP), The Ohio State University, Columbus, OH 43210, USA}

\author{Tuan Dung Nguyen}
\affiliation{Department of Computer and Information Science, University of Pennsylvania, Philadelphia, PA 19104, USA}

\author{Tirthankar Ghosal}
\affiliation{National Center for Computational Sciences, Oak Ridge National Laboratory, Oak Ridge, TN 37831, USA}

\author{Rui Pan (潘瑞）}
\affiliation{Department of Computer Science and Engineering, Hong Kong University of Science and Technology, Kowloon, Hong Kong}

\author{Hardik Arora}
\affiliation{Indian Institute of Technology Patna, Bihta, Bihar 801106, India}

\author{Zechang Sun (孙泽昌）}
\affiliation{Department of Astronomy, MongManWai Building, Tsinghua University, Beijing 100084, China}

\author{Tijmen de Haan}
\affiliation{Institute of Particle and Nuclear Studies, High Energy Accelerator Research Organization, Tsukuba, Ibaraki 305-0801, Japan}
\affiliation{International Center for Quantum-field Measurement Systems for Studies of the Universe and Particles (QUP-WPI), High Energy Accelerator Research Organization (KEK), Tsukuba, Ibaraki 305-0801, Japan}

\author{Nesar Ramachandra}
\author{Azton Wells}
\affiliation{Computational Science Division, Argonne National Laboratory, Lemont, IL 60439, USA}

\author{Sandeep Madireddy}
\affiliation{Mathematics and Computer Science Division, Argonne National Laboratory, Lemont, IL 60439, USA}

\author{Alberto Accomazzi}
\affiliation{Center for Astrophysics, Harvard \& Smithsonian, Cambridge, MA 02138, USA}

\begin{abstract}
We present a comprehensive evaluation of proprietary and open-weights large language models using the first astronomy-specific benchmarking dataset. This dataset comprises 4,425 multiple-choice questions curated from the Annual Review of Astronomy and Astrophysics, covering a broad range of astrophysical topics.\footnote{Jeopardy is a popular American quiz show where contestants are tested on their knowledge across various subjects. Other similar shows include Who's Still Standing (一站到底) in China and University Challenge in the UK, among others.} Our analysis examines model performance across various astronomical subfields and assesses response calibration, crucial for potential deployment in research environments. Claude-3.5-Sonnet outperforms competitors by up to 4.6 percentage points, achieving 85.0\% accuracy. For proprietary models, we observed a universal reduction in cost every 3-to-12 months to achieve similar score in this particular astronomy benchmark. open-weights models have rapidly improved, with LLaMA-3-70b (80.6\%) and Qwen-2-72b (77.7\%) now competing with some of the best proprietary models. We identify performance variations across topics, with non-English-focused models generally struggling more in exoplanet-related fields, stellar astrophysics, and instrumentation related questions. These challenges likely stem from less abundant training data, limited historical context, and rapid recent developments in these areas. This pattern is observed across both open-weights and proprietary models, with regional dependencies evident, highlighting the impact of training data diversity on model performance in specialized scientific domains. Top-performing models demonstrate well-calibrated confidence, with correlations above 0.9 between confidence and correctness, though they tend to be slightly underconfident. The development for fast, low-cost inference of open-weights models presents new opportunities for affordable deployment in astronomy. The rapid progress observed suggests that LLM-driven research in astronomy may become feasible in the near future.
\end{abstract}


\section{Introduction}

The emergence of the GPT (Generative Pre-trained Transformers) model series has thrust large language models (LLMs) into the spotlight \citep{brown2020language, kaplan2020}, showcasing their diverse capabilities in language comprehension and reasoning \citep{Nelson2019, achiam2023gpt}. These advancements hold the potential to revolutionize astronomical research methodologies. As astronomy has expanded, individual subfields have become increasingly isolated due to the vast literature requiring assimilation. LLMs' extensive abilities could prove crucial in developing more robust recommender systems \citep{Geng2022,Chu2023,Zhao2023,Vats2024} to aid human researchers and powerful tools to summarize field evolution through knowledge graphs \citep{Kau2024,Sun2024}, thereby inspiring future research. Moreover, the emerging demonstrable reasoning ability of LLMs offers the possibility of deploying them as research agents \citep{Boiko2023,Bran2023,Ramos2024}, enabling the automation of individual downstream tasks or even facilitating end-to-end research. This could allow for the analysis and reasoning of myriad cosmic sources \citep{Laureijs2011,Aihara2018} that would otherwise only receive cursory examination under current hand-crafted and human-driven analyses.

For both recommender systems and research agents, a critical ability lies in the LLM's capacity to understand modern astronomical contexts \citep{Beltagy2019}, both in terms of knowledge recall and of deriving robust inferences based on the latest consensus of the astronomical research community. While numerous routine tests have been deployed to benchmark LLMs \citep{gao2023,Zheng2023,Chiang2024,Zheng2024}, these assessments, though somewhat representative, remain tangential to an LLM agent's ability to perform astronomical research. This is partly due to the nature of astronomical research, which often requires broad logical thinking and the connection of different knowledge domains \citep{Chen2018}. Research in astronomy frequently necessitates drawing inspiration from cross-domain knowledge while keeping pace with both technological and statistical developments \citep{Hu2002,Abbott2016}.

Several established benchmarks have been designed to evaluate different aspects of model performance, such as MMLU \citep{hendrycks2020measuring}, BIG-bench \citep{srivastava2023Beyond}, HELM \citep{liang2022holistic}, SuperGLUE \citep{wang2019superglue}, and TruthfulQA \citep{lin2021truthfulqa}. While these benchmarks are valuable for assessing general language understanding and reasoning capabilities, they fall short in evaluating the specific skills required for astronomical research. Most of these tests cover a broad range of subjects but lack a specific focus on astronomy or fail to capture the depth of knowledge required in the field. The absence of precise benchmarks that test the broad knowledge of LLMs in astronomical research, or more generally in scientific Q\&A, remains a key shortcoming in the current development and benchmarking of LLMs \citep{Yasunaga2019,Luo2022,Saikh2022}. This limitation partly stems from the prohibitive cost, if not impossibility, of creating human-annotated benchmarking datasets in PhD-level scientific research domains \citep{Bowman2015}.

However, such benchmarks are critical for two main reasons. Firstly, some LLMs risk being overtrained on specific, well-developed benchmarks, which might not be representative of their generalizability when deployed as research agents in astronomy. Secondly, as we will demonstrate, even some of the more well-known proprietary and open-weights models can vary drastically in their performance on astronomical benchmarks, equivalent to three orders of magnitude in cost inefficiency even when they perform equally well on more established benchmarks like MMLU. As much of the potential for deploying LLM agents relies on understanding cost-efficient calculations, a better benchmarking system that establishes a baseline is urgently needed, especially given that more than 1.5 years has passed since the groundbreaking announcement of ChatGPT.

In this study, we address this gap by providing a comprehensive evaluation of various proprietary and open-weights LLMs in the astronomical context. Our work establishes a robust benchmark that accurately assesses the capabilities of LLMs in astronomical research, particularly their ability to recall astronomical facts and make broad inferences based on current astronomical consensus. This benchmark aims to facilitate more informed decisions in LLM deployment and further development. 

We note that while Retrieval-Augmented Generation (RAG) techniques could potentially enhance LLM performance in retrieving astronomical information, we deliberately focus on testing the native capabilities of these models without RAG implementation. This decision stems from several considerations. Effective RAG deployment involves complex choices in text embeddings, fine-tuning approaches, and retrieval strategies that warrant dedicated investigation. Moreover, raw text often proves insufficient for RAG, necessitating sophisticated summarization techniques. These complexities merit a separate paper specifically examining optimal RAG strategies for astronomical applications.

This paper is the first in a series of related studies. Subsequent papers will release more specific details about the curation of our benchmarking datasets, including detailed Q\&A, contrast the performance of baseline models with in-house continually pre-trained astronomy-specific LLMs, and conduct detailed arena battles for base models with further fine-tuned versions. Through this comprehensive series of studies, we aim to establish a new standard for evaluating LLMs in the context of astronomy and pave the way for more effective integration of AI technologies in astronomical research.

Our paper is structured as follows: We begin with a brief discussion on the curation of the multiple-choice question (MCQ) datasets used in this benchmark and the inference details for LLM-based answering. We then present a comparison of proprietary LLM models' performance, from earlier iterations to recent developments. Subsequently, we contrast the performance of proprietary models with open-weights alternatives. We discuss the implications of our findings for the future of LLMs in astronomical research and conclude with a summary of key findings and their significance.

\vspace{2cm}
\section{Benchmarking MCQ Datasets}

As part of the initiative in collaboration with the Argonne National Laboratory, we have developed detailed astronomical benchmarking datasets comprising both Q\&A and MCQ components. These datasets are specifically designed to evaluate the performance of LLMs in the context of astronomical research, a topic we will explore in greater detail in our subsequent paper. Our benchmark not only tests astronomical facts and consensus from the research community but also assesses models' capabilities in linking insights across diverse subfields and understanding the interdisciplinary nature of astronomical research.

Traditionally, creating such datasets has been hindered by the high cost of human annotation, particularly in specialized scientific domains. However, astrophysics and astronomy benefit from a long-standing tradition of world-leading experts summarizing the state of the field. The Annual Review of Astronomy and Astrophysics stands out as an invaluable resource in this regard. Established in 1963, this review journal, with an impact factor of 33.3 in 2023, publishes approximately ten articles annually. Reviews are commissioned by invitation from a panel of senior and prominent members of the editorial committee. The highly selective nature of the journal ensures that each article provides an overview of cutting-edge research in a specific subfield of astronomy, typically spanning an average of 40 pages and 15,000 words. This approach precludes any myopic views on particular topics, and the contributing authors are widely recognized as world leaders in their respective fields.

The extensive length of these reviews initially posed challenges for models with shorter context windows. However, recent advancements in long-context LLMs have made it possible to extract quality MCQs from these reviews. When this study began, Gemini-1.5-Pro~\citep{team2023gemini} offered the longest context window of one million tokens, and was widely available via Google's Generative AI API. While Gemini-1.5-Pro is not the most performant in terms of offline astronomical Q\&A, we deemed it sufficient for generating this dataset because it can digest entire articles in its context. We employed extensive prompt engineering to ensure the quality of the MCQs, which will be summarized below and described in more detail in the second paper of this series.

To create this Q\&A dataset, we started with collecting 885 articles in ARAA, dating from 1963 to 2023. Then, we used the Nougat optical character recognition (OCR) tool~\citep{blecher2023nougat} to transcribe these papers into text. We fed each paper into Gemini-1.5-Pro and instructed it to propose 5 questions that can be answered based on the paper's content. Each question was accompanied by four options (A, B, C, D) only one of which is correct. We prompted the model with three important instructions: (1) make the question specific enough to the article's content but general enough so that it can later be asked independently; (2) also make the answer general enough, e.g., by refraining from pointing to specific sections of the article; and (3) ensure that the four answers are about the same in length, which stems from our earlier review that the model tends to propose the correct answer to be the longest. Additionally, the model was tasked with providing an explanation and cited paragraphs from the review supporting the answer. This process yielded a total of 4,425 questions. Below are some of the examples (and see more examples in Appendix~\ref{appendixC} and ~\ref{appendixD}
    
\begin{questionbox}[boxcolor2]

    \textbf{Paper ID:} 2023ARA\&A..61..373F\\

    \textbf{Question:} What is the primary reason for the decline in the number density of luminous quasars at redshifts greater than 5?
    
    (A) A decrease in the overall star formation rate, leading to fewer potential host galaxies for quasars.
    
    (B) An increase in the neutral hydrogen fraction in the intergalactic medium, which obscures the quasars' light.
    
    (C) A decrease in the number of massive black hole seeds that can form and grow into supermassive black holes.
    
    (D) An increase in the average metallicity of the Universe, leading to a decrease in the efficiency of black hole accretion.\\

    \textbf{Correct Answer:} C \\

    \textbf{Explanation:} The article discusses how the number density of luminous quasars decreases exponentially at redshifts greater than 5, suggesting that the earliest quasars emerge at a redshift of approximately 10. This decline is attributed to a decrease in the availability of massive black hole seeds, which are necessary for the formation and growth of supermassive black holes that power quasars. As the Universe ages and expands, the conditions for the formation of these massive seeds become less favorable, leading to a decrease in the number of quasars.
\end{questionbox}

\begin{questionbox}[boxcolor3]
    \textbf{Paper ID:} 2023ARA\&A..61..473C \\

    \textbf{Question:} What is the primary goal of calibrating subgrid feedback models in cosmological simulations? 
    
    (A) To ensure that simulations accurately reproduce the observed properties of the interstellar medium.
    
    (B) To create a diverse range of galaxy morphologies in the simulations.
    
    (C) To achieve convergence in simulation results across different resolutions and box sizes.
    
    (D) To steer simulations towards producing a broadly realistic galaxy population that is consistent with key observational constraints. \\

    \textbf{Correct Answer:} D \\

    \textbf{Explanation:} The calibration of subgrid feedback models is primarily done to ensure that simulations produce a galaxy population that broadly aligns with key observational constraints. This is crucial because the microphysics governing feedback processes occur on scales much smaller than the resolution of cosmological simulations. By calibrating these models, simulations can better reproduce properties like the galaxy stellar mass function and the relationship between galaxy stellar mass and central supermassive black hole mass. This is discussed in section 2.4 of the article.
\end{questionbox}

\begin{questionbox}[boxcolor1]
    \textbf{Paper ID:} 2023ARA\&A..61..131F\\
    
    \textbf{Question:} The properties of the circumgalactic medium (CGM) primarily depend on the competition between:
        
    (A) Star formation rate and supernova feedback.
        
    (B) Gas cooling and stellar winds.
        
    (C) Gravity-driven infall and gas cooling.
        
    (D) Magnetic fields and thermal conduction.\\
        
    \textbf{Correct Answer:} C\\
        
    \textbf{Explanation:} The article explicitly states that the defining characteristic of the CGM is the balance between gravity pulling gas inwards and cooling processes that allow gas to lose pressure and condense. This balance dictates whether the CGM is predominantly hot (slow cooling) or cold (rapid cooling).
\end{questionbox}

Most questions pertain to key domain knowledge in the field, which is unsurprising given that Annual Reviews often serve as references for expert astronomers. Due to careful prompt engineering, the questions are general enough to be answered as standalone queries, without referring to results from specific research articles. Human experts reviewed a subset of these examples and, based on the provided explanations, deemed the quality adequate.

We acknowledge that this automated process of MCQ generation using LLMs is a compromise \citep[see also][]{Zhang2024}. The high cost of human expertise made it impractical to hand-craft such a large number of high-quality, doctoral-level benchmarks. Consequently, the accuracy we present is certainly a lower limit, as some answers might be inaccurate—for instance, questions pertaining to older review articles may contain outdated information. Additionally, some questions might be too vague to provide accurate answers. However, as we will demonstrate, strong LLMs show robust accuracy performance (up to 85\%). More importantly, the performance roughly aligns with the average strength of LLM models generally, indicating that the vast majority of questions are (a) challenging enough to trip up weaker models and (b) overall answerable with ground truth results. We found that limiting our analysis to questions accurately answered by the strongest model (a strong indicator of correct answers) would not significantly alter our conclusions.

The details of these MCQ benchmarks will be presented in a forthcoming companion paper. In particular, we will focus on our on-going efforts to validate these proposed Q\&As, i.e., ensuring that the questions and accompanying answers are correct and sufficiently difficult. For now, we contend that these models provide a sufficiently robust benchmarking dataset. Interestingly, while the questions were generated using Gemini, our results show that Gemini performs worse than other equivalent models. This demonstrates that the questions curated by Gemini-1.5 do not visibly favor Gemini models when their performance is tested.

\section{Inference Methodology}

For this study, we focus exclusively on the MCQ benchmarks, reserving more detailed evaluations of astronomical research capabilities, such as open-ended Q\&A, which were also curated as part of our effort, for future work. The latter may require more careful human curation through an interactive platform to ensure meaningful results.

In testing the MCQ performance, particularly for open-weights models, we faced several methodological choices. One option was to use base models—LLMs primarily trained on next-token prediction without specialized fine-tuning (SFT) for question-answering alignment. This approach, used in other benchmarking studies \citep{Plaut2024}, involves a computationally efficient method where, after presenting the question, the prompt ``The answer is" is given, and the logit (probability) of the next token being A, B, C, or D is calculated.

However, we opted for a more computationally intensive method that we believe offers greater accuracy and insight. We used the Instruct/Chat versions of the models (see Appendix~\ref{appendixA} for prompt details) and requested both an answer and an explanation. This decision was motivated by two key factors: (1) SFT alignment in earlier and weaker (especially earlier open-weights) models might not consistently follow direct instructions. For instance, the tokens following ``The answer is" might include variations like ``A" or ``The answer is, in my opinion," among other combinations. As a result, less performant models where the logits for all options (A to D) can be orders of magnitude lower than one, unlike in more capable models such as those with 70B parameters.

(2) Although, we noticed that open-weights base models without SFT can occasionally show slight improvements over instruct models when using the first-token approach. However, such performance does not translate well to practical deployment as LLM research agents, which typically require multi-turn reasoning and more nuanced instruction following. This insight inspired us to primarily test the performance with the Instruct/Chat models instead of the base models.

In our evaluation of both proprietary and open-weights models, we adhered to the default instructions (e.g., temperature and top-p settings) provided in the API documentation for proprietary models or the Hugging Face instructions for open-weights models. We acknowledge that individual models may have room for improvement through model-specific prompt fine-tuning or hyperparameter optimization. However, we have observed that, especially for robust recent models, setting the temperature of the inference, even at extreme values of 1 (most stochastic output) and 0 (zero stochasticity), often leads to nearly identical scores. This is likely due to the extensive specialized fine-tuning of alignments that these models undergo. We consider such optimizations beyond the scope of this study, which aims to provide an overview of existing models in the most generic and unbiased manner practically possible. Our goal is to offer a bird's-eye view of model performance under standard conditions, rather than pushing the limits of each model's capabilities through extensive customization.

In our prompt (see Appendix~\ref{appendixA}), we found that chain-of-thought prompting, which requires models to consider the reasoning behind their rationale, generally improves the accuracy of most models in answering MCQs \citep{Wei2022,Zhou2022}.  Our decision is supported by observations that LLMs are implicit thinkers, and prompting them to provide an explanation helps improve accuracy, a finding corroborated by our results.

For weaker models, particularly earlier open-weights versions, adherence to the specified JSON output format was inconsistent. We implemented a preliminary regex to extract answers, which was successful in nearly all cases (approximately 99\% of the time). For the remaining instances where regex failed, we passed the answer and explanation through a GPT-4o model to determine the intended answer. This approach further demonstrates the utility of including explanations in the output.

In rare cases (typically $<$1\% for any given model), some models opted not to answer certain questions, either deeming them unreliable or unable to provide a single definitive answer. Handling these cases presents a challenge: it could indicate that the model is strong enough to identify flaws in the MCQ questions, or conversely, that it's too weak to formulate an answer. To maintain fairness across all models, we decided to discard, on an individual basis, questions that models refused to answer. While this introduces some variability between models, such cases are infrequent. For stronger models, this affected at most 0.2\% of questions (10 out of 4,425). If anything, this approach may benefit weaker models that refuse to answer more frequently. Importantly, this decision does not significantly impact the overall benchmarking results presented in this study.
\begin{figure*}
\centering
\includegraphics[width=1.0\textwidth]{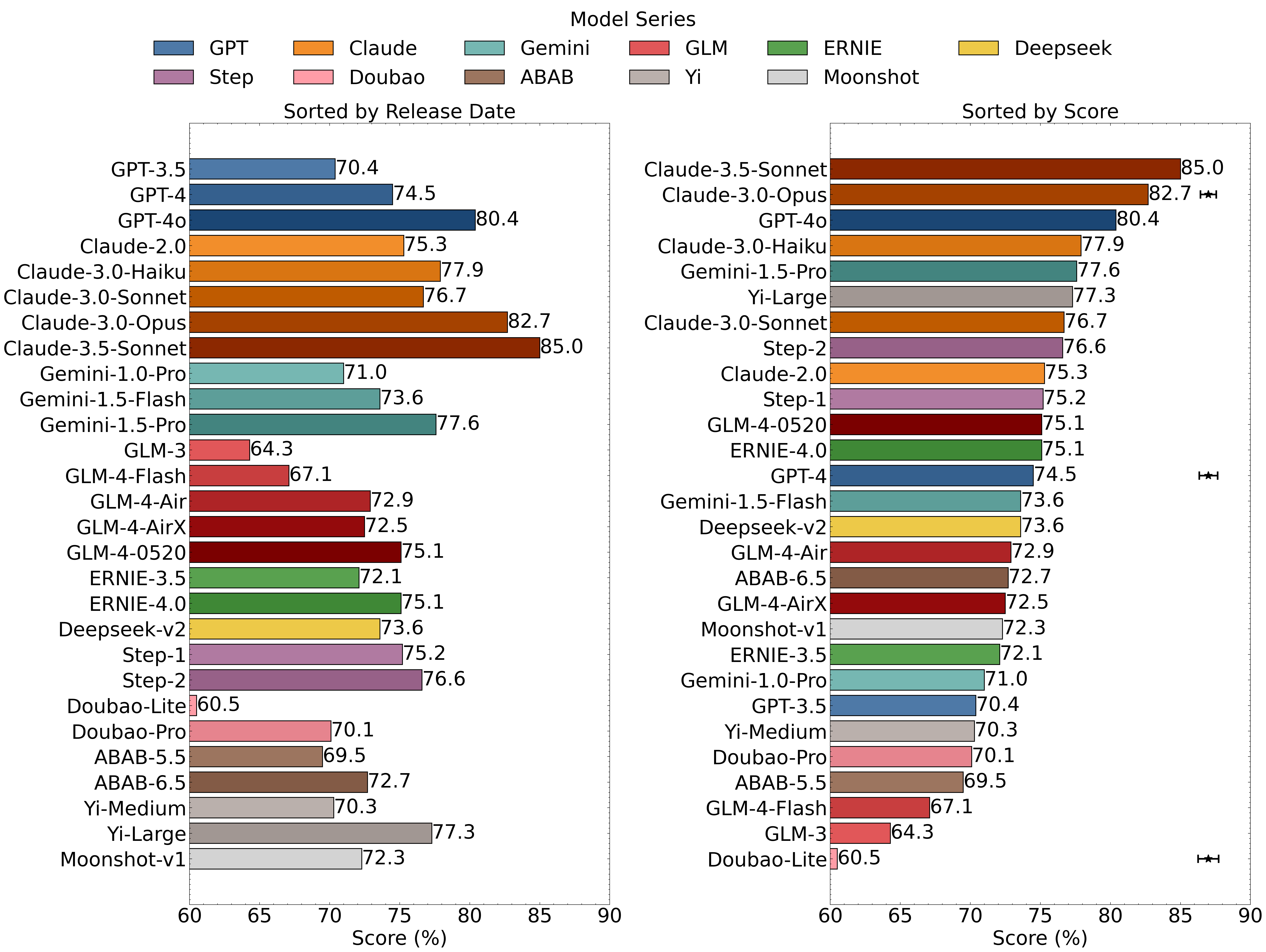}
\caption{Benchmarking scores of proprietary large language models for MCQ answering in astronomical research. The left panel groups models by series, with darker shades indicating more recent or larger models within each series. We tested GPT-3.5, GPT-4, GPT-4o, Claude-2.0, Claude-3.0 (Haiku, Sonnet, Opus), Claude-3.5-Sonnet, Gemini-1.0-Pro, Gemini-1.5 (Flash, Pro), GLM-3, GLM-4 (Flash, Air, AirX, 0520), Ernie-3.5, Ernie-4.0, Deepseek-v2, Step-1, Step-2,  Doubao (Lite, Pro), ABAB-5.5, ABAB-6.5, Yi (Medium, Large), and Moonshot-v1. Claude-3.5-Sonnet performs best with an 85.0\% accuracy, outperforming the closest non-Anthropic competitor, GPT-4o, by 4.6 percentage points. Among other leading models, GLM-4-0520 achieves 75.1\%, showing a gap of 9.9 percentage points from the top performer. Interestingly, while many cutting-edge models perform similarly in general benchmarks, they show significant variability in this niche astronomical research question-answering task. The performance gap can be as large as 14.9 percentage points (between Claude-3.5-Sonnet and Doubao-Pro), demonstrating the need for domain-specific benchmarks. The right panel shows the same scores sorted by overall performance, regardless of model series, highlighting the wide range of capabilities across different models in this specific task. The error bars in the right panel display the Wilson Score Interval as uncertainties ($\pm 0.6-0.8$ percentage points) for three representative models, reflecting the statistical variation due to the finite set of 4,425 questions.}
\label{fig1}
\end{figure*}

\begin{figure*}
\centering
\includegraphics[width=1.0\textwidth]{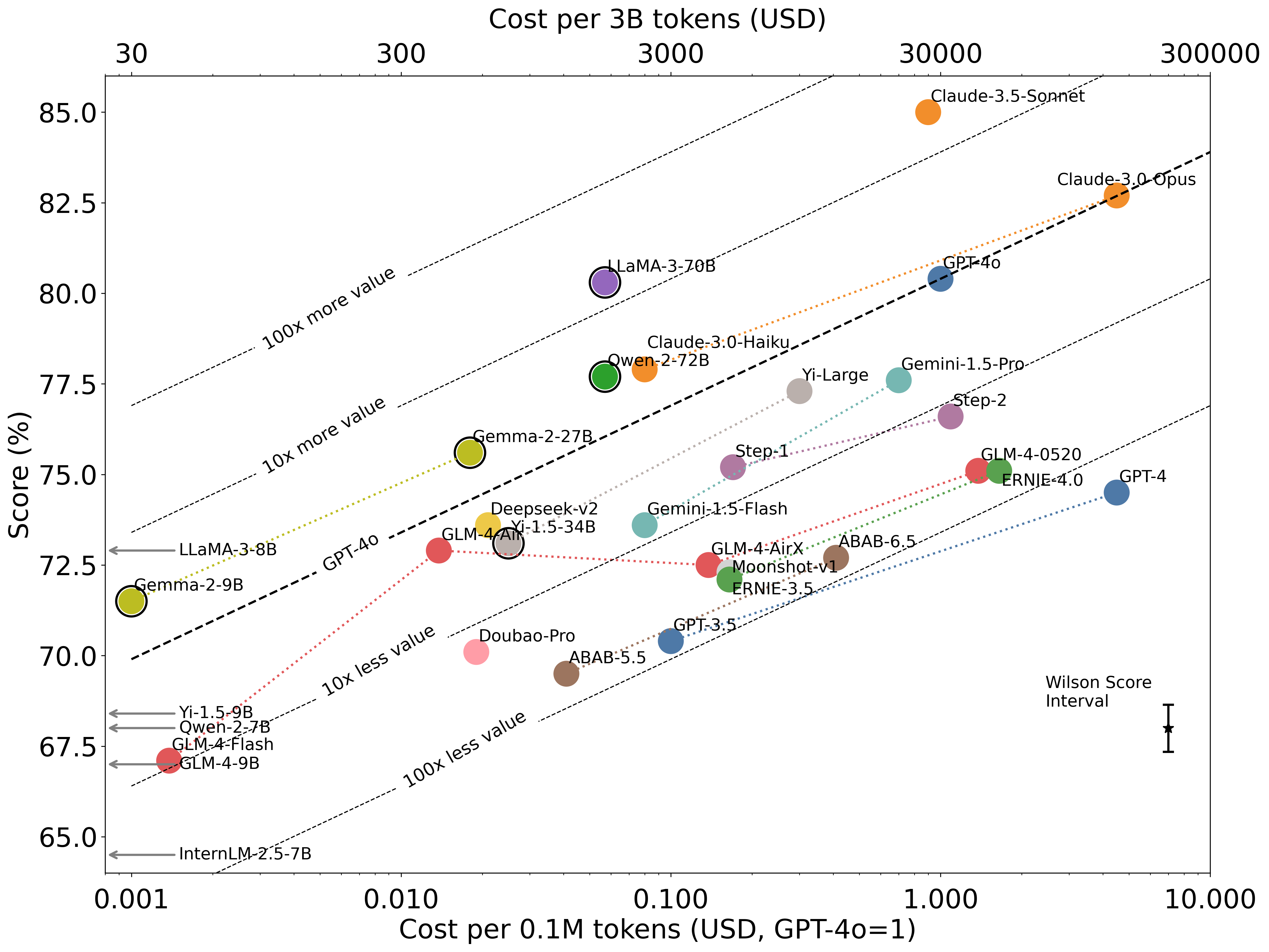}
\caption{Cost and performance trade-off in astronomical Q\&A. The dual x-axes show the cost per 0.1 million tokens (typical for agent deployment on one astronomical source; see text for details) and the cost to process an ArXiv astro-ph worth of tokens ($\sim$3B tokens). We use the average of input and output token costs based on June 2024 prices. To avoid crowding, only representative models are shown; full performances are in Tables~\ref{table1} and \ref{table2}. Models with an outer circle indicate open-weights models run on low-cost APIs, leveraging recent specialized GPU developments for transformers. Left arrows indicate cheaper open-weights models below the plot's lower bound. Dotted lines of the same color connect models of the same series. Generally, within a series, there's a 10-fold cost increase for a 3.5-point accuracy improvement. Dashed guidelines show equivalent performance accounting for cost trade-offs, with the bold dashed line showing GPT-4o's value. Claude-3.5-Sonnet outperforms others models. LLaMA-3-70B is the only model in the same tier, albeit with lower performance. Second-tier models include Gemman-2-9B, Gemma-2-27B, Qwen-2-72B, Claude-3.0-Haiku, performing similarly to GPT-4o and Claude-3.0-Opus when price is considered. However, for GPT-4-like performance ($>$80\% accuracy), only Claude-3.5-Sonnet, Claude-3.0-Opus, and LLaMA-3-70B qualify. A representative the Wilson Score Interval as uncertainty, calculated for a 75\% accuracy rate over the 4,425 questions, is displayed in the bottom right corner for reference.}
\label{fig2}
\end{figure*}

\section{Benchmarking Proprietary Large Language Models}

We begin by presenting the performance of proprietary models. The accuracy metric used throughout is defined as the percentage of questions answered correctly. Given our dataset of 4,425 questions, the typical noise, evaluated with the Wilson Score Interval in our evaluation is small ($1\sigma$ range is about $\pm 0.6-0.8\%$). 

The Wilson Score Interval \citep{Wilson1927} provides a confidence interval for binomial proportions that outperforms the normal approximation, particularly for extreme probabilities and small sample sizes. Given a sample of size $n$ with $k$ successes and sample proportion $\hat{p} = k/n$, the Wilson score interval for the true population proportion $p$ is:
$$
\frac{\hat{p} + \frac{z^2}{2n} \pm z\sqrt{\frac{\hat{p}(1-\hat{p})}{n} + \frac{z^2}{4n^2}}}{1 + \frac{z^2}{n}}
$$
\noindent
Here, $z$ is the $(1+\text{confidence level})/2$ quantile of the standard normal distribution. For a 1$\sigma$ interval, we have $z = 1$ exactly. As $N$ increases or for non-extreme values of $p$, this interval converges to the Poisson noise estimate, making it a robust choice across various scenarios.

Consequently, most differences between models that we will discuss are statistically significant. Our analysis focuses on several leading proprietary series that we consider representative of the current state-of-the-art:

\begin{enumerate}
\item OpenAI's GPT series (GPT-3.5, GPT-4, GPT-4o)\footnote{\url{https://openai.com/index/openai-api}}

\item Anthropic's Claude series (Claude-2.0, Claude-3.0 Haiku/Sonnet/Opus, Claude-3.5-Sonnet)\footnote{\url{https://www.anthropic.com/api}}

\item Google's Gemini series (Gemini-1.0-Pro, Gemini-1.5-Flash/Pro)\footnote{\url{https://ai.google.dev/gemini-api}}

\item ZhiPu's GLM series (GLM-3-Turbo, GLM-4 Flash/Air/AirX/0520)\footnote{\url{https://open.bigmodel.cn/dev/api}}

\item Baidu's ERNIE series (ERNIE-3.5, ERNIE-4.0)\footnote{\url{https://qianfan.cloud.baidu.com/}}

\item Deepseek's series (Deepseek-v2)\footnote{\url{https://platform.deepseek.com/}}

\item Stepfun's series (Step-1, Step-2)\footnote{\url{https://platform.stepfun.com/}}

\item ByteDance's Doubao series (Doubao-Lite, Doubao-Pro)\footnote{\url{https://www.volcengine.com/docs/82379/1263512}}

\item MiniMax AI's ABAB series (ABAB-5.5, ABAB-6.5)\footnote{\url{https://www.minimaxi.com/}}

\item 01.AI's Yi series (Yi-Medium, Yi-Large)\footnote{\url{https://platform.01.ai/}}

\item Moonshot AI's Kimi series (Moonshot-v1)\footnote{\url{https://platform.moonshot.cn/}}
\end{enumerate}

This list, while not exhaustive, encompasses most competitive proprietary models. We acknowledge the existence of other competitive proprietary models. As we progress towards releasing our benchmarking dataset following more rigorous human evaluation, we encourage model developers to contact our group for benchmarking opportunities. 

We note that, by definition, the lowest score a model could potentially achieve is about 25\% from random guessing. We decided not to normalize the scores based on this baseline to keep the accuracy more literal and meaningful on an absolute scale.

\begin{table*}[htbp]
\centering
\caption{Performance and Pricing of Various Proprietary Language Models on Astronomy MCQ Benchmarking. The first column show the score in percentage and the other two columns show the price to run the proprietary models by processing 0.1M and 3B tokens. Pricing reflects average input and output costs as of June 2024, with applicable exchange rates. Cost per 0.1M tokens is based on typical usage for an LLM agent performing simple agentic tasks. The 3B token represents roughly the amount of words in the entire arXiv astro-ph archive (till March 2024), approximating the token count for the complete astronomy literature available on astro-ph. In the first column, the best three proprietary models are highlighted with gold, silver, and bronze symbols, and the score of the best model is emboldened.}
\label{table1}
\begin{tabular}{lccc}
\hline
\textbf{Model} & \textbf{Score (\%)} & \textbf{Cost per 0.1M tokens (in USD)} & \textbf{Cost per 3B tokens (in USD)} \\
\hline
\multicolumn{4}{l}{\textbf{OpenAI/GPT Series}} \\
GPT-3.5 & 70.4 & \$0.10 & \$3,000 \\
GPT-4 & 74.5 & \$4.50 & \$135,000 \\
GPT-4o & 80.4 \bronzemedal & \$1.00 & \$30,000 \\[0.1cm]
\hline
\multicolumn{4}{l}{\textbf{Anthropic/Claude Series}} \\[0.1cm]
Claude-2.0 & 75.3 & \$1.60 & \$48,000 \\
Claude-3.0-Haiku & 77.9 & \$0.08 & \$2,400 \\
Claude-3.0-Sonnet & 76.7 & \$0.90 & \$27,000 \\
Claude-3.0-Opus & 82.7 \silvermedal & \$4.50 & \$135,000 \\[0.1cm]
Claude-3.5-Sonnet & \textbf{85.0} \goldmedal & \$0.90 & \$27,000 \\[0.1cm]
\hline
\multicolumn{4}{l}{\textbf{Google/Gemini Series}} \\
Gemini-1.0-Pro & 71.0 & \$0.10 & \$3,000 \\[0.1cm]
Gemini-1.5-Flash & 73.6 & \$0.08 & \$2,400 \\
Gemini-1.5-Pro & 77.6 & \$0.70 & \$21,000 \\[0.1cm]
\hline
\multicolumn{4}{l}{\textbf{Zhipu(智谱)/GLM Series}} \\
GLM-3-Turbo & 64.3 & \$0.014 & \$420 \\[0.1cm]
GLM-4-Flash & 67.1 & \$0.0014 & \$42 \\
GLM-4-Air & 72.9 & \$0.014 & \$420 \\
GLM-4-AirX & 72.5 & \$0.138 & \$4,140 \\
GLM-4-0520 & 75.1 & \$1.38 & \$41,400 \\[0.1cm]
\hline
\multicolumn{4}{l}{\textbf{Baidu/ERNIE(文心一言) Series}} \\
ERNIE-3.5 & 72.1 & \$0.165 & \$4,950 \\
ERNIE-4.0 & 75.1 & \$1.65 & \$49,500 \\[0.1cm]
\hline
\multicolumn{4}{l}{\textbf{Deepseek(深度求索) Series}} \\
Deepseek-v2 & 73.6 & \$0.021 & \$630 \\[0.1cm]
\hline
\multicolumn{4}{l}{\textbf{Step(阶跃星辰) Series}} \\
Step-1 & 75.2 & \$0.17 & \$5,100 \\
Step-2 & 76.6 & \$1.09 & \$32,700\\[0.1cm]
\hline
\multicolumn{4}{l}{\textbf{ByteDance/Doubao(豆包) Series}} \\
Doubao-Lite & 60.5 & \$0.006 & \$180 \\
Doubao-Pro & 70.1 & \$0.019 & \$570 \\[0.1cm]
\hline
\multicolumn{4}{l}{\textbf{MiniMax AI Series}} \\
ABAB-5.5 & 69.5 & \$0.041 & \$1,230 \\
ABAB-6.5 & 72.7 & \$0.41 & \$12,300 \\[0.1cm]
\hline
\multicolumn{4}{l}{\textbf{01/Yi(零一万物) Series}} \\
Yi-Medium & 70.3 & \$0.034 & \$1,020 \\
Yi-Large & 77.3 & \$0.30 & \$9,000 \\[0.1cm]
\hline
\multicolumn{4}{l}{\textbf{Moonshot(月之暗面)/Kimi Series}} \\
Moonshot-v1 & 72.3 & \$0.165 & \$4,950 \\[0.1cm]
\hline
\end{tabular}
\end{table*}

\subsection{Overall Performance Comparison Across Model Series}

We first study the overall performance of the models over the entire 4,425 questions. The histogram in Fig.\ref{fig1} provides an overview of the results. The left panel groups models by series, with darker shades indicating more recent or larger models within each series. The right panel shows the same scores sorted by overall performance, regardless of model series. These results are summarized in Table~\ref{table1} (the corresponding results for open-weights model, which we will discuss in Section~\ref{sec:open-models}, can be found in Table~\ref{table2}). The overall results range from Doubao-Lite at 60.5\% to Claude-3.5-Sonnet at 85.0\%. These results demonstrate that while the MCQ questions are challenging (see example questions in Table~\ref{table1}, Appendiex~\ref{appendixC} and \ref{appendixD}), they are manageable for LLMs, with overall performance roughly aligning with each model's general strength based on other benchmarks. It's particularly noteworthy that LLMs can handle such a specialized domain, given that astronomy is a niche research topic (with about 3B tokens\footnote{AstroMLab ran OCR on the entire astro-ph arXiv archive from 1993 to 2024. We found it contains roughly 2.6 billion words, which here our estimate translates to about 3 billion tokens. While this is admittedly a rough calculation, it provides a ballpark figure for the costs involved when dealing with this volume of text} in ArXiv astro-ph out of the trillions of tokens used to train these models) with a relatively small community of active practitioners compared to broader fields like medical science.

However, even among the latest proprietary models with arguably similar reasoning abilities (e.g., as gauged by MMLU), performance varies by almost twenty-five points. For instance, while released around the same time (June 2024), Doubao-Pro scores around 70.1\%, while Claude-3.5-Sonnet achieves 85.0\%. Even among popular models, the results can be vastly different: Claude-3.5-Sonnet achieves 85.0\%, Gemini-1.5-Pro reaches 77.6\%, and GPT-4o scores 80.4\%, highlighting the importance of this benchmark. It demonstrates that general model capabilities do not always translate into robust performance in astronomy-specific tasks.

Nonetheless, as showin in the figure, unsurprisingly, within a given class of models, later versions generally outperform earlier ones. For example, Claude 2.0 achieves 75.3\% accuracy, which improves to 82.7\% in Claude-3.0-Opus and then to 85.0\% with Claude-3.5-Sonnet. Similarly, GPT-3.5 scores 70.4\%, while GPT-4 and GPT-4o achieve 74.5\% and 80.4\% respectively. For Gemini, we see an improvement from 71.0\% for Gemini-1.0-Pro to 77.6\% for Gemini-1.5-Pro. Low-cost models often perform substantially worse than their more expensive counterparts; GLM-4-Flash achieves 67.1\%, while the top-performing GLM series model, GLM-4-0520, reaches 75.1\%. Similarly, Yi-Medium scores 70.3\%, while Yi-Large achieves 77.3\%.

Interestingly, we observe some variation in performance among models developed in different regions. Among the models that are not primarily focused on English-language content, despite often performing equally well in the general benchmark (notably, e.g., \citealt{GLM2024}), appears to fall slightly behind in our particular benchmark, the highest-performing model from this group is Yi-Large \citep{01AI2024}, with 77.3\% accuracy, which is notably strong but still 7.7 percentage points behind the top-performing model. This variation could potentially be attributed to differences in training data, focus areas during model development, or other factors specific to astronomy as a domain. We will further explore this in Section~\ref{sec:why-weak}, and Section~\ref{sec:temporal}

A human researcher reviewed dozens of questions that even Claude-3.5-Sonnet answered incorrectly, judging the answers based on the explanations and context extracted during MCQ generation. While some problematic questions were identified, an issue that we will address in the second paper of this series for a more detailed curation of our benchmarks, most appeared accurate to expert eyes, indicating significant room for improvement for LLMs to reach the level of an all-knowing astronomy expert.

\begin{figure*}
\centering
\includegraphics[width=1\textwidth]{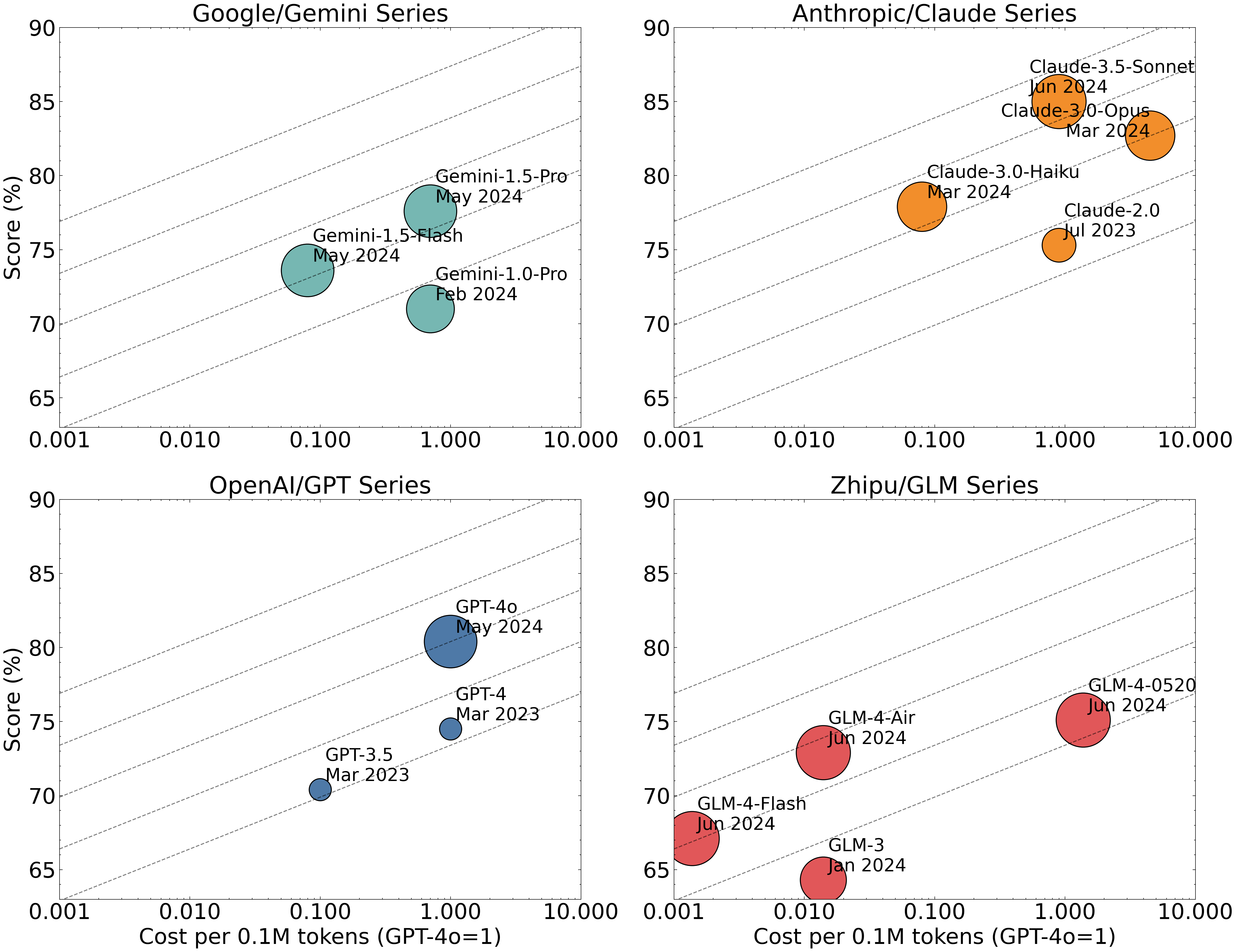}
\caption{The Cost Efficiency Improvement Rate for Proprietary Models. This figure demonstrates the trade-off between astronomical MCQ answering accuracy and price for representative examples of proprietary models that have released multiple series: OpenAI/GPT, Anthropic/Claude, Google/Gemini, and Zhipu/GLM. The size of each point represents the recency of the model's release, with larger points indicating more recent releases. The dashed lines, similar to Fig.~\ref{fig2}, show improvements in cost-efficiency, where moving up one line represents a 3.5-point increase in score for the same cost, or equivalently, a 10x improvement in value for the same performance (see text for details). For all these models, we observe rapid improvements in performance and cost-efficiency over time: Gemini improved equivalent to about a 100x improvement in cost-efficiency in three months (Gemini-1.0 to Gemini-1.5). Claude progressed to a 10x improvement in cost-efficiency over about 3 months (Claude-3.0 to Claude-3.5). The GPT series improved by 30x in cost efficiency over about 14 months (GPT-3.5 and GPT-4 to GPT-4o). The GLM series shows improvements of about 10-100x in cost efficiency within 6 months (GLM-3 to GLM-4).}
\label{fig3}
\end{figure*}

\subsection{Cost Efficiency Consideration}

While proprietary models vary in ability, a key consideration for deploying LLM agents is cost-effectiveness. In the following analysis, we consider the pricing of these models as of June 2024, when this draft is being written. We focus on the cost per 0.1M (0.1 million) tokens as a basic unit for the cost. This metric is chosen based on a companion work (Sun et al., in prep.) where we deploy LLM agents for astronomical research. We find that a multi-turn conversation of 0.1M tokens is typically needed for reasoning about individual astronomical sources. This number multiplies with the number of sources for such studies, providing an interesting rule of thumb for management considerations in future deployments of LLMs for active astronomical research.

For many proprietary models, the input and output prices differ. Deploying LLMs often requires both extensive input tokens (e.g., for retrieval-augmented generation in the context of custom generative models for various coding APIs) and output tokens for reasoning, multi-turn conversations, and collaborative agent efforts. Given these considerations, we will use the average of the input and output token prices in our analysis, assuming they play an equal role. As a reference point, GPT-4o, perhaps the most well-known LLM, costs exactly 1 US dollar per 0.1M tokens, which further informed our decision to use this as our benchmarking unit for the cost.

Fig.~\ref{fig2} shows the score of various representative proprietary LLMs plotted against their cost. For a complete list of scores for all models tested, refer to Table~\ref{table1}. First, within a given model series, we observe a universal scaling across different model series, which provides an interesting insight into the trade-off between performance and cost. Our analysis reveals that most proprietary models, within a fixed series at a given publication time, follow a universal trade-off of approximately 10 times cost increase per 3.5-point score improvement. These are shown in dashed lines that link the models within series, such as GPT-3.5 and GPT-4, Gemini-3.0-Haiku and Gemini-3.0-Opus, Gemini-1.5-Pro and Gemini-1.5-Flash, the GLM-4 series (0520, AirX, Air, Flash), ERNIE-4.0 and ERNIE-3.5, and ABAB-5.5 and ABAB-6.5. And this universal tradeoff is even more pronounced when we only restrict to questions extracted from Annual Reviews post-1990 (see Appendix~\ref{appendixB}).

This relationship can be approximated by the equation:
\begin{equation}
{\rm Score} (\%) = 3.5 \log_{10}({\rm Cost_{normalized}}) + 80
\end{equation}

\noindent
where ${\rm Cost_{normalized}}$ is the cost relative to GPT-4o (i.e., GPT-4o has a normalized cost of 1, with a score of 80.4\%). This equation signifies that we are trading off about 10 times the relative cost for an equivalent of 3.5 points of improvement in score.

In Fig.~\ref{fig2}, we plot reference lines based on this relationship. The bold dashed line represents the case where we have an intercept of 80.4\% at the normalized cost of 1 (equivalent to GPT-4o's cost). Additional dashed lines with the same slope are offset by 3.5 points in score, where each offset signifies an equivalent of paying an order of magnitude extra for the same performance.

This universal scaling showcases a key insight: for tasks such as astronomical research recall and summarization - including creating knowledge graphs and deploying LLM agents - the cost for a desired performance can vary by more than three orders of magnitude.  In fact, between GPT-3.5 and 4 versus the latest GPT-4o, this benchmark has revealed a two order of magnitude improvement in this trade-off. Indeed, the old GPT-3.5 and 4, while sometimes still widely applied in the API for casual users, are amongst the worst in this metric. This demonstrates the importance of such benchmarking for the development of LLM in astronomy.

Following these tilted guide lines, which represent equal cost-effectiveness at different performance levels, \emph{among the proprietary models} (i.e., ignoring the models with black outer circles), Claude-3.5-Sonnet is the obvious winner as of June 2024. Claude-3.0-Haiku, GPT-4o, and Claude-3.0-Opus, while varying 56-fold in cost, have trade-offs that make them equally desirable models after Claude-3.5-Sonnet. Beyond that, Yi-Large, GLM-4-Air and DeepSeek-v2 are also competitive, with DeepSeek-v2 and GLM-4-Air, providing the most affordable price point. In fact, with GLM-4-Air's price point of USD 0.014 per 0.1M tokens, dealing with 3B tokens (roughly the size of astro-ph) would cost only USD 420. Gemini-1.5, despite its prominence in other benchmarks, falls slightly behind the above models.

Some of the models including Doubao-Pro from ByteDance, GLM from Zhipu AI \citep{GLM2024}, Alibaba's ERNIE \citep{Ernie2021}, and ABAB from MiniMax AI, despite their impressive abilities in other benchmarks, appear to fall short in our benchmarks. Finally, we refrain from discussing the open-weights models in Fig~\ref{fig2}, including LLaMA-3, Qwen-2, and Gemma-2, and will return to comparing these models in Section~\ref{sec:open-models}.

\vspace{2cm}

\subsection{How Fast is the Cost Efficiency Improving for Astronomical Tasks?}

A key consideration for deploying LLM agents for various astronomical tasks, apart from their overall reasoning ability and robust knowledge recall and summarization, comes down to cost considerations. For instance, while GLM-4-Flash can process the entire astronomy arXiv for less than USD 42 (as of June 2024), it's the accuracy of the model that might make the most difference. This is particularly important for scientific research, where the accuracy of recall is often non-negotiable. Therefore, at a targeted desired performance level, how quickly the price improves will critically determine if LLM research agents can be deployed at scale.

Fig.~\ref{fig3} provides a more quantitative assessment of this rate across four major model series: Google's Gemini, Anthropic's Claude, OpenAI's GPT, and Zhipu's GLM. The size of each point represents the recency of the model's initial release, with larger points indicating more recent releases. The dashed lines show improvements in cost-efficiency, where moving up one line represents a 3.5-point increase in score for the same cost, or equivalently, a 10x reduction in cost for the same performance.

All panels paint a consistent picture: on average, at a desired performance level, the pricing is improving by about an order of magnitude every 3-12 months. Specifically:

\begin{enumerate}
\item Google's Gemini series improved from 71.0\% to 77.6\% accuracy in just three months (Gemini-1.0-Pro in February 2024 to Gemini-1.5-Pro in May 2024) at the same cost point, equivalent to about a 100x improvement in cost-efficiency.
\item Anthropic's Claude series showed remarkable progress. It improved from 75.3\% (Claude-2.0, July 2023) to 77.9\% (Claude-3.0-Haiku, March 2024) with Claude-3.0-Haiku 10 times lower the price yet performing visibly better than Claude-2.0. Then, it further improved to 85.0\% (Claude-3.5-Sonnet, June 2024) in just another 3 months, representing another order of magnitude leap in cost-efficiency. In total, this represents about a 1000x improvement in price-efficiency from Claude-2.0 to Claude-3.5-Sonnet over 11 months, or 10x from Claude-3.0 to Claude-3.5 over 3 months.
\item OpenAI's GPT series improved from 74.5\% (GPT-4, March 2024) to 80.4\% (GPT-4o, May 2024) fourteen month later. The latter represents about a 30x improvement in cost-efficiency. OpenAI, perhaps unsurprisingly given its pioneering work, exhibits a slightly slower growth compared to the other companies. 
\item Zhipu's GLM series shows improvements from 64.3\% (GLM-3, January 2024) to 72.9\% (GLM-4-Air, June 2024) at the same cost point within six months, while also offering higher-performance options like GLM-4-0520 at 75.1\%. This represents about a 10-100x improvement in cost-efficiency from GLM-3 to GLM-4. 
\end{enumerate}

These improvements point to a period required for a 10-fold increase in cost-efficiency of 3-12 months across different model series. However, we stress that this trend should be viewed as a rough guideline rather than a hard rule. Our cost-price analysis assumes the current performance-price relationship based on the pricing of various proprietary models (see the dotted lines in Fig.~\ref{fig2}). It's crucial to note that different companies have their own priorities (such as multimodality or context window length), and a model's release date often differs significantly from its training cut-off date. We chose to focus on release dates, as training cut-off information isn't consistently available.

\subsection{Why Are the Weaker Proprietary Models Weaker?}
\label{sec:why-weak}

While the overall accuracy metric provides a comprehensive view of each model's ability in astronomical knowledge recall, dissecting the differences between models in detail can lead to insights about the major trade-offs when opting for a cheaper model or models that appear to be slightly over-tuned for other metrics, resulting in performance decreases in this particular benchmarking.

To this end, for individual questions, we relied on GPT-4o to perform two classifications. For the first classification, we categorized the questions by topics, following the different subclasses in astro-ph articles: (1) Solar and Stellar Astrophysics, (2) Earth and Planetary Astrophysics, (3) Astrophysics of Galaxies, (4) Cosmology and Nongalactic Astrophysics, (5) High Energy Astrophysics, and (6) Instrumentation and Methods for Astrophysics. For the second evaluation, we asked GPT-4o to further classify questions into the following groups based on the different abilities being tested: (1) Understanding Fundamental Concepts, (2) Technical and Observational Techniques, (3) Analytical and Reasoning Skills, (4) Historical and Theoretical Knowledge, and (5) Current Research and Advanced Topics.

\begin{table*}[htbp]
\centering
\caption{Performance of Open-Weights Language Models on Astronomy MCQ Benchmarks. The first column shows the average accuracy of each model. The top three models are indicated with a gold symbol and two silver symbols (due to a tie for second place), and the highest score is in bold. In the subsequent four columns, we compare the scores with four representative proprietary models (Claude-3.5-Sonnet, GPT-4o, Gemini-1.5-Pro, GPT-3.5). Scores of open-weights models that exceed proprietary benchmarks are bolded.}
\label{table2}
\begin{tabular}{lccccc}
\hline
\textbf{Model} & \textbf{Score (\%)} & \textbf{Δ Claude-3.5-Sonnet (\%)} & \textbf{Δ GPT-4o (\%)} & \textbf{Δ Gemini-1.5-Pro (\%)} & \textbf{Δ GPT-3.5 (\%)} \\
\hline
\multicolumn{6}{l}{\textbf{Meta/LLaMA Series}} \\
LLaMA-2-7B & 50.3 & -35.0 & -30.1 & -27.3 & -20.1 \\
LLaMA-2-70B & 70.7 & -14.6 & -9.7 & -6.9 & \textbf{+0.3} \\[0.1cm]
LLaMA-3-8B & 72.9 & -12.4 & -7.5 & -4.7 & \textbf{+2.5} \\
LLaMA-3-70B & \textbf{80.6} \goldmedal & -4.4 & \textbf{+0.2} & \textbf{+3.0} & \textbf{+10.2} \\[0.1cm]
\hline
\multicolumn{6}{l}{\textbf{Mistral AI Series}} \\
Mistral-7B-v0.1 & 48.1 & -36.9 & -32.0 & -29.5 & -22.3 \\
Mistral-8x7B-v0.1 & 73.7 & -11.3 & -6.4 & -3.9 & \textbf{+3.3} \\
Mixtral-8x22B-v0.1 & 77.7 \silvermedal & -7.3 & -2.4 & \textbf{+0.1} & \textbf{+7.3} \\[0.1cm]
Mistral-7B-v0.2 & 62.1 & -22.9 & -18.0 & -15.5 & -8.3 \\[0.1cm]
Mistral-7B-v0.3 & 63.9 & -21.1 & -16.2 & -13.7 & -6.5 \\[0.1cm]
\hline
\multicolumn{6}{l}{\textbf{Microsoft/Phi Series}} \\
Phi-2-3B & 65.6 & -19.4 & -14.5 & -12.0 & -4.8 \\[0.1cm]
Phi-3-4B & 71.7 & -13.3 & -8.4 & -5.9 & \textbf{+1.3} \\
Phi-3-14B & 75.6 & -9.4 & -4.5 & -2.0 & \textbf{+5.2} \\[0.1cm]
\hline
\multicolumn{6}{l}{\textbf{Google/Gemma Series}} \\
Gemma-1-2B & 44.1 & -40.9 & -36.0 & -33.5 & -26.3 \\
Gemma-1-7B & 56.1 & -28.9 & -24.0 & -21.5 & -14.3 \\[0.1cm]
Gemma-2-9B & 71.5 & -13.5 & -8.6 & -6.1 & \textbf{+1.1} \\
Gemma-2-27B & 75.6 & -9.4 & -4.5 & -2.0 & \textbf{+5.2} \\[0.1cm]
\hline
\multicolumn{6}{l}{\textbf{Alibaba/Qwen(通义千问) Series}} \\
Qwen-1-7B & 57.4 & -27.6 & -22.7 & -20.2 & -13.0 \\[0.1cm]
Qwen-1.5-7B & 63.7 & -21.3 & -16.4 & -13.9 & -6.7 \\
Qwen-1.5-14B & 67.7 & -17.3 & -12.4 & -9.9 & -2.7 \\
Qwen-1.5-32B & 73.2 & -11.8 & -6.9 & -4.4 & \textbf{+2.8} \\
Qwen-1.5-110B & 72.7 & -12.3 & -7.4 & -4.9 & \textbf{+2.3} \\[0.1cm]
Qwen-2-7B & 68.0 & -17.0 & -12.1 & -9.6 & -2.4 \\
Qwen-2-57B & 71.8 & -13.2 & -8.3 & -5.8 & \textbf{+1.4} \\
Qwen-2-70B & 77.7 \silvermedal & -7.3 & -2.4 & \textbf{+0.1} & \textbf{+7.3} \\[0.1cm]
\hline
\multicolumn{6}{l}{\textbf{01/Yi(零一万物) Series}} \\
Yi-1.5-6B & 61.0 & -24.0 & -19.1 & -16.6 & -9.4 \\
Yi-1.5-9B & 68.4 & -16.6 & -11.7 & -9.2 & -2.0 \\
Yi-1.5-34B & 73.1 & -11.9 & -7.0 & -4.5 & \textbf{+2.7} \\[0.1cm]
\hline
\multicolumn{6}{l}{\textbf{Deepseek(深度求索) Series}} \\
Deepseek-67B & 63.1 & -21.9 & -17.0 & -14.5 & -7.3 \\[0.1cm]
\hline
\multicolumn{6}{l}{\textbf{Zhipu(智谱)/ChatGLM Series}} \\
ChatGLM3-6B & 50.4 & -34.6 & -29.7 & -27.2 & -20.0 \\
GLM-4-9B & 67.0 & -18.0 & -13.1 & -10.6 & -3.4 \\[0.1cm]
\hline
\multicolumn{6}{l}{\textbf{PJ Lab(浦语)/InternLM(书生) Series}} \\
InternLM-2.5-7B & 64.5 & -20.5 & -15.6 & -13.1 & -5.9 \\[0.1cm]
\hline
\end{tabular}
\end{table*}

\begin{table*}[htbp]
\centering
\small
\caption{Performance of Proprietary Large Language Models on Astronomy Multiple Choice Questions by Subfield Topic. Scores are presented as percentages, indicating the fraction of correctly answered MCQs. Models are grouped by series. For each topic, the highest score within the proprietary models is in bold. Including open-weights models from Table~\ref{table4}, we award gold, silver, and bronze medals to the top three performers per topic (unless scores are tied). Blue medals indicate honorable mentions, ranking 4th-6th.}
\label{table3}
\begin{tabular}{lcccccc}
\hline
\textbf{Model} & \textbf{Solar \&} & \textbf{Earth \&} & \textbf{Galactic} & \textbf{Cosmology \&} & \textbf{High Energy} & \textbf{Instrumentation} \\
 & \textbf{Stellar} & \textbf{Planetary} & \textbf{Astrophysics} & \textbf{Nongalactic} & \textbf{Astrophysics} & \textbf{\& Methods} \\
\hline
\multicolumn{7}{l}{\textbf{OpenAI/GPT Series}} \\
GPT-3.5 & 67.9 & 69.6 & 70.6 & 75.3 & 73.9 & 71.6 \\
GPT-4 & 73.0 & 73.5 & 76.3 & 73.3 & 77.3 & 73.7 \\
GPT-4o & 78.4 \notable & 80.7 \bronzemedal & 82.0 \notable & 81.9 \bronzemedal & 82.4 \notable & 79.3 \notable \\[0.1cm]
\hline
\multicolumn{7}{l}{\textbf{Anthropic/Claude Series}} \\
Claude-2.0 & 74.1 & 72.4 & 77.8 & 73.1 & 76.9 & 76.6 \\[0.1cm]
Claude-3.0-Haiku & 77.6 \notable & 74.3 & 79.3 \notable & 78.2 \notable & 80.9 \notable & 75.4 \\
Claude-3.0-Sonnet & 76.1 \notable & 73.7 & 77.6 & 76.4 & 79.9 & 76.3 \\
Claude-3.0-Opus & 81.5 \silvermedal & 83.9 \silvermedal & 83.6 \silvermedal & \textbf{84.1} \goldmedal & 83.2 \bronzemedal & 81.9 \silvermedal \\[0.1cm]
Claude-3.5-Sonnet & \textbf{83.6} \goldmedal & \textbf{85.2} \goldmedal & \textbf{87.0} \goldmedal & 83.0 \silvermedal & \textbf{85.5} \goldmedal & \textbf{84.6} \goldmedal \\[0.1cm]
\hline
\multicolumn{7}{l}{\textbf{Google/Gemini Series}} \\
Gemini-1.0-Pro & 69.5 & 71.3 & 70.2 & 73.4 & 74.7 & 71.3 \\[0.1cm]
Gemini-1.5-Flash & 72.2 & 72.6 & 74.3 & 74.8 & 77.4 & 71.3 \\
Gemini-1.5-Pro & 75.4 & 78.0 \notable & 78.5 & 75.9 & 80.8  & 78.4 \notable \\[0.1cm]
\hline
\multicolumn{7}{l}{\textbf{Zhipu(智谱)/GLM Series}} \\
GLM-3-Turbo & 62.8 & 62.4 & 63.9 & 70.1 & 66.9 & 66.3\\[0.1cm]  
GLM-4-Flash & 64.7 & 67.4 & 66.9 & 70.8 & 70.0 & 69.5 \\
GLM-4-Air & 71.0 & 71.5 & 73.7 & 73.8 & 76.5 & 72.8 \\
GLM-4-AirX & 70.0 & 70.7 & 73.2 & 74.9 & 77.0 & 73.7 \\
GLM-4-0520 & 72.4 & 75.0 & 76.1 & 77.4 & 78.6 & 75.7 \\[0.1cm]
\hline
\multicolumn{7}{l}{\textbf{Baidu/ERNIE(文心一言) Series}} \\
ERNIE-3.5 & 70.7 & 72.0 & 72.1 & 77.5 & 75.1 & 68.9 \\
ERNIE-4.0 & 73.6 & 72.8 & 76.5 & 76.7 & 78.4 & 73.1 \\[0.1cm]
\hline
\multicolumn{7}{l}{\textbf{01/Yi(零一万物) Series}} \\
Yi-Medium & 67.9 & 68.3 & 72.7 & 72.7 & 75.1 & 64.5 \\
Yi-Large & 75.6 & 75.9 & 78.9 & 76.0 & 79.6 & 76.3 \\[0.1cm]
\hline
\multicolumn{7}{l}{\textbf{Deepseek(深度求索) Series}} \\
Deepseek-v2 & 71.5 & 71.5 & 75.0 & 71.6 & 77.8 & 72.8 \\[0.1cm]
\hline
\multicolumn{7}{l}{\textbf{Step(阶跃星辰) Series}} \\
Step-1 & 72.7 & 74.3 & 78.0 & 74.2 & 79.2 & 71.0 \\
Step-2 & 74.3 & 75.7 & 78.2 & 76.8 \notable & 81.1 \notable & 73.7 \\[0.1cm]
\hline
\multicolumn{7}{l}{\textbf{ByteDance/Doubao(豆包) Series}} \\
Doubao-Lite & 58.3 & 62.0 & 59.6 & 65.3 & 61.5 & 66.0 \\
Doubao-Pro & 68.1 & 68.0 & 71.3 & 71.2 & 75.1 & 68.0 \\[0.1cm]
\hline
\multicolumn{7}{l}{\textbf{MiniMax AI/ABAB Series}} \\
ABAB-5.5 & 67.3 & 67.8 & 70.6 & 70.5 & 71.1 & 72.2 \\
ABAB-6.5 & 70.3 & 71.1 & 74.1 & 74.9 & 76.7 & 71.3 \\[0.1cm]
\hline
\multicolumn{7}{l}{\textbf{Moonshot(月之暗面)/Kimi Series}} \\
Moonshot-v1 & 70.4 & 70.9 & 72.7 & 75.7 & 75.6 & 70.9 \\[0.1cm]
\hline
\end{tabular}
\end{table*}

\begin{table*}[htbp]
\centering
\small
\caption{Performance of Open-Weights Large Language Models on Astronomy Multiple Choice Questions by Subfield Topic. This table follows the same format as Table~\ref{table3}, but for open-weights models. Scores are presented as percentages, indicating the fraction of correctly answered MCQs. Models are grouped by series. For each topic, the highest score within the open-weights models is in bold. Combined with proprietary models from Table~\ref{table3}, we award gold, silver, and bronze medals to the top three performers per topic (unless scores are tied). Blue medals indicate honorable mentions, ranking 4th-6th.}
\label{table4}
\begin{tabular}{lcccccc}
\hline
\textbf{Model} & \textbf{Solar \&} & \textbf{Earth \&} & \textbf{Galactic} & \textbf{Cosmology \&} & \textbf{High Energy} & \textbf{Instrumentation} \\
 & \textbf{Stellar} & \textbf{Planetary} & \textbf{Astrophysics} & \textbf{Nongalactic} & \textbf{Astrophysics} & \textbf{\& Methods} \\
\hline
\multicolumn{7}{l}{\textbf{Meta/LLaMA Series}} \\
LLaMA-2-7B & 49.4 & 46.7 & 51.7 & 48.3 & 52.1 & 52.1 \\
LLaMA-2-70B & 69.1 & 65.7 & 72.4 & 73.4 & 71.9 & 74.0 \\[0.1cm]
LLaMA-3-8B & 71.2 & 70.7 & 73.2 & 74.2 & 76.5 & 75.1 \\
LLaMA-3-70B & \textbf{78.7} \bronzemedal & \textbf{79.1} \notable & \textbf{82.1} \bronzemedal & \textbf{81.5} \notable & \textbf{83.4} \silvermedal & \textbf{80.2} \bronzemedal \\[0.1cm]
\hline
\multicolumn{7}{l}{\textbf{Mistral AI Series}} \\
Mistral-7B-v0.1 & 47.0 & 44.1 & 47.7 & 54.8 & 49.3 & 53.0 \\
Mistral-8x7B-v0.1 & 72.1 & 73.9 & 73.7 & 76.8 & 78.1 & 71.3 \\
Mixtral-8x22B-v0.1 & 75.8 & 77.6 & 80.0 \notable & 77.8 & 80.7 & 75.7 \\[0.1cm]
Mistral-7B-v0.2 & 59.0 & 58.5 & 63.6 & 64.9 & 65.9 & 66.9 \\[0.1cm]
Mistral-7B-v0.3 & 62.6 & 60.7 & 64.7 & 64.9 & 66.2 & 65.4 \\[0.1cm]
\hline
\multicolumn{7}{l}{\textbf{Microsoft/Phi Series}} \\
Phi-2-3B & 65.0 & 63.1 & 66.3 & 67.7 & 63.2 & 70.7 \\[0.1cm]
Phi-3-4B & 70.7 & 72.2 & 71.0 & 71.2 & 74.0 & 73.1 \\
Phi-3-14B & 74.0 & 75.7 & 75.6 & 74.5 & 77.7 & 78.4 \notable \\[0.1cm]
\hline
\multicolumn{7}{l}{\textbf{Google/Gemma Series}} \\
Gemma-1-2B & 42.8 & 42.0 & 44.9 & 51.7 & 42.9 & 46.7 \\
Gemma-1-7B & 54.0 & 53.9 & 55.6 & 60.1 & 59.2 & 60.7 \\[0.1cm]
Gemma-2-9B & 71.0 & 70.4 & 70.6 & 70.8 & 74.6 & 72.5 \\
Gemma-2-27B & 73.3 & 77.8 \notable & 77.6 & 75.6 & 76.8 & 73.4 \\[0.1cm]
\hline
\multicolumn{7}{l}{\textbf{Alibaba/Qwen(通义千问) Series}} \\
Qwen-1-7B & 56.3 & 58.3 & 56.8 & 64.6 & 57.8 & 58.0 \\[0.1cm]
Qwen-1.5-7B & 62.5 & 58.7 & 63.1 & 69.4 & 68.4 & 63.9 \\
Qwen-1.5-14B & 67.1 & 62.6 & 68.0 & 72.8 & 72.2 & 63.5 \\
Qwen-1.5-32B & 70.6 & 69.1 & 75.5 & 72.1 & 78.8 & 73.1 \\
Qwen-1.5-110B & 71.2 & 71.5 & 73.3 & 71.6 & 76.5 & 72.2 \\[0.1cm]
Qwen-2-7B & 66.1 & 68.9 & 67.4 & 67.2 & 71.5 & 71.3 \\
Qwen-2-57B & 69.1 & 69.6 & 73.5 & 74.6 & 75.3 & 71.1 \\
Qwen-2-70B & 76.1 \notable & 76.5 & 78.5 & 78.6 \notable & 80.0 & 78.6 \notable \\[0.1cm]
\hline
\multicolumn{7}{l}{\textbf{01/Yi(零一万物) Series}} \\
Yi-1.5-6B & 59.3 & 61.0 & 61.4 & 67.5 & 61.7 & 60.2 \\
Yi-1.5-9B & 66.6 & 64.3 & 70.4 & 68.3 & 70.8 & 69.8 \\
Yi-1.5-34B & 70.3 & 68.0 & 75.8 & 74.2 & 77.8 & 73.1 \\[0.1cm]
\hline
\multicolumn{7}{l}{\textbf{Deepseek(深度求索) Series}} \\
Deepseek-67B & 60.4 & 63.8 & 63.4 & 65.5 & 67.9 & 61.9 \\[0.1cm]
\hline
\multicolumn{7}{l}{\textbf{Zhipu(智谱)/ChatGLM Series}} \\
ChatGLM3-6B & 49.7 & 48.6 & 50.2 & 55.1 & 50.8 & 51.0 \\
GLM-4-9B & 65.0 & 65.4 & 67.1 & 67.5 & 71.3 & 69.2 \\[0.1cm]
\hline
\multicolumn{6}{l}{\textbf{PJ Lab(浦语)/InternLM(书生) Series}} \\
InternLM-2.5-7B & 62.5 & 63.0 & 65.7 & 62.7 & 69.6 & 63.9\\[0.1cm]
\hline
\end{tabular}
\end{table*}

\begin{table*}[htbp]
\centering
\small
\caption{Performance of Proprietary Large Language Models on Astronomy Multiple Choice Questions by Tested Ability. This table follows a similar format to Table~\ref{table3}, but groups questions based on the abilities they test rather than subfield topics. Scores are presented as percentages, indicating the fraction of correctly answered MCQs. Models are grouped by series. For each ability, the highest score within the proprietary models is in bold. Combined with open-weights models from Table~\ref{table6}, we award gold, silver, and bronze medals to the top three performers per ability (unless scores are tied). Blue medals indicate honorable mentions, ranking 4th-6th.}
\label{table5}
\begin{tabular}{lccccc}
\hline
\textbf{Model} & \textbf{Understanding} & \textbf{Technical \&} & \textbf{Analytical \&} & \textbf{Historical \&} & \textbf{Advanced} \\
 & \textbf{Fundamental} & \textbf{Observational} & \textbf{Reasoning} & \textbf{Theoretical} & \textbf{Topics} \\
  & \textbf{Concepts} & \textbf{Techniques} & \textbf{Skills} & \textbf{Knowledge} & \\
\hline
\multicolumn{6}{l}{\textbf{OpenAI/GPT Series}} \\
GPT-3.5 & 78.5 & 74.2 & 66.7 & 67.5 & 74.1 \\
GPT-4 & 77.8 & 77.3 & 69.4 & 71.3 & 79.5 \\
GPT-4o & 82.5 \notable & 81.0 \notable & 73.6 & 78.9 \notable & 88.3 \silvermedal \\[0.1cm]
\hline
\multicolumn{6}{l}{\textbf{Anthropic/Claude Series}} \\
Claude-2.0 & 78.5 & 77.0 & 70.8 & 71.1 & 77.2 \\[0.1cm]
Claude-3.0-Haiku & 81.1 & 77.1 & 69.4 & 75.9 & 81.5 \\
Claude-3.0-Sonnet & 81.7 \notable & 78.5 & 73.6 & 73.5 & 79.6 \\
Claude-3.0-Opus & 84.8 \silvermedal & 82.1 \silvermedal & 76.4 \notable & 82.5 \silvermedal & 87.0 \bronzemedal \\[0.1cm]
Claude-3.5-Sonnet & \textbf{86.4} \goldmedal & \textbf{86.2} \goldmedal & 77.8 \silvermedal & \textbf{84.3} \goldmedal & \textbf{89.4} \goldmedal \\[0.1cm]
\hline
\multicolumn{6}{l}{\textbf{Google/Gemini Series}} \\
Gemini-1.0-Pro & 77.3 & 73.3 & 61.1 & 71.1 & 69.8 \\[0.1cm]
Gemini-1.5-Flash & 78.9 & 75.2 & 63.9 & 68.1 & 74.5 \\
Gemini-1.5-Pro & 81.1 & 78.1 & 73.6 & 75.9 & 81.4 \\[0.1cm]
\hline
\multicolumn{6}{l}{\textbf{Zhipu(智谱)/GLM Series}} \\
GLM-3-Turbo & 71.3 & 67.1 & 65.3 & 65.7 & 66.0 \\[0.1cm]
GLM-4-Flash & 73.4 & 70.0 & 62.5 & 62.7 & 71.6 \\
GLM-4-Air & 76.8 & 75.0 & 59.7 & 73.5 & 79.6 \\
GLM-4-AirX & 76.1 & 75.2 & 56.9 & 69.3 & 80.9 \\
GLM-4-0520 & 79.1 & 76.0 & 69.4 & 75.0 & 77.6 \\[0.1cm]
\hline
\multicolumn{6}{l}{\textbf{Baidu/ERNIE(文心一言) Series}} \\
ERNIE-3.5 & 77.8 & 73.5 & 72.2 & 72.3 & 76.5 \\
ERNIE-4.0 & 80.5 & 76.7 & 72.2 & 72.9 & 80.3 \\[0.1cm]
\hline
\multicolumn{6}{l}{\textbf{01/Yi(零一万物) Series}} \\
Yi-Medium & 76.0 & 71.5 & 62.5 & 65.1 & 75.9 \\
Yi-Large & 79.5 & 80.0 \notable & 70.8 & 72.9 & 84.0 \notable \\[0.1cm]
\hline
\multicolumn{6}{l}{\textbf{Deepseek(深度求索) Series}} \\
Deepseek-v2 & 78.5 & 76.7 & 72.2 & 70.5 & 76.5 \\[0.1cm]
\hline
\multicolumn{6}{l}{\textbf{Step(阶跃星辰) Series}} \\
Step-1 & 79.9 & 73.6 & \textbf{79.2} \goldmedal & 75.9 & 82.7 \\
Step-2 & 79.9 & 77.3 & 77.7 \bronzemedal & 72.9 & 83.3 \\[0.1cm]
\hline
\multicolumn{6}{l}{\textbf{ByteDance/Doubao(豆包) Series}} \\
Doubao-Lite & 67.5 & 62.8 & 63.9 & 58.4 & 66.1 \\
Doubao-Pro & 76.4 & 71.9 & 66.7 & 66.3 & 69.8 \\[0.1cm]
\hline
\multicolumn{6}{l}{\textbf{MiniMax AI/ABAB Series}} \\
ABAB-5.5 & 72.8 & 72.7 & 65.3 & 66.3 & 76.5 \\
ABAB-6.5 & 78.0 & 73.5 & 69.4 & 68.1 & 72.8 \\[0.1cm]
\hline
\multicolumn{6}{l}{\textbf{Moonshot(月之暗面)/Kimi Series}} \\
Moonshot-v1 & 77.0 & 75.2 & 65.3 & 70.3 & 78.3 \\[0.1cm]
\hline
\end{tabular}
\end{table*}

\begin{table*}[htbp]
\centering
\small
\caption{Performance of Open-Weights Large Language Models on Astronomy Multiple Choice Questions by Tested Ability. This table follows the same format as Table~\ref{table5}, but for open-weights models. Scores are presented as percentages, indicating the fraction of correctly answered MCQs. Models are grouped by series. For each ability, the highest score within the open-weights models is in bold. Combined with proprietary models from Table~\ref{table5}, we award gold, silver, and bronze medals to the top three performers per ability (unless scores are tied). Blue medals indicate honorable mentions, ranking 4th-6th.}
\label{table6}
\begin{tabular}{lccccc}
\hline
\textbf{Model} & \textbf{Understanding} & \textbf{Technical \&} & \textbf{Analytical \&} & \textbf{Historical \&} & \textbf{Advanced} \\
 & \textbf{Fundamental} & \textbf{Observational} & \textbf{Reasoning} & \textbf{Theoretical} & \textbf{Topics} \\
  & \textbf{Concepts} & \textbf{Techniques} & \textbf{Skills} & \textbf{Knowledge} & \\
\hline
\multicolumn{6}{l}{\textbf{Meta/LLaMA Series}} \\
LLaMA-2-7B & 54.5 & 51.4 & 44.4 & 54.2 & 50.6 \\
LLaMA-2-70B & 76.4 & 72.7 & 61.1 & 66.9 & 74.1 \\[0.1cm]
LLaMA-3-8B & 78.0 & 75.0 & 65.3 & 75.3 & 74.7 \\
LLaMA-3-70B & \textbf{84.1} \bronzemedal & \textbf{82.1} \silvermedal & 73.6 & 78.9 \notable & 84.0 \notable \\[0.1cm]
\hline
\multicolumn{6}{l}{\textbf{Mistral AI Series}} \\
Mistral-7B-v0.1 & 54.1 & 55.4 & 47.2 & 53.0 & 57.4 \\
Mistral-8x7B-v0.1 & 78.3 & 75.4 & 68.1 & 69.3 & 80.9 \\
Mixtral-8x22B-v0.3 & 81.7 \notable & 77.8 & \textbf{76.4} \notable & 74.6 & 85.6 \notable \\[0.1cm]
Mistral-7B-v0.2 & 67.1 & 66.3 & 58.3 & 61.4 & 64.8 \\[0.1cm]
Mistral-7B-v0.3 & 69.3 & 66.1 & 66.7 & 62.7 & 66.1 \\[0.1cm]
\hline
\multicolumn{6}{l}{\textbf{Microsoft/Phi Series}} \\
Phi-2-3B & 70.1 & 70.1 & 60.0 & 69.8 & 66.9 \\[0.1cm]
Phi-3-4B & 76.0 & 71.9 & 66.7 & 75.3 & 73.5 \\
Phi-3-14B & 79.3 & 76.4 & 68.1 & \textbf{80.1} \bronzemedal & 79.0 \\[0.1cm]
\hline
\multicolumn{6}{l}{\textbf{Google/Gemma Series}} \\
Gemma-1-2B & 47.1 & 48.5 & 43.1 & 44.0 & 50.0 \\
Gemma-1-7B & 60.4 & 59.5 & 50.0 & 60.2 & 54.9 \\[0.1cm]
Gemma-2-9B & 75.4 & 71.9 & 69.4 & 70.5 & 74.1 \\
Gemma-2-27B & 78.1 & 75.6 & 72.2 & 77.7 \notable & 80.9 \\[0.1cm]
\hline
\multicolumn{6}{l}{\textbf{Alibaba/Qwen(通义千问) Series}} \\
Qwen-1-7B & 63.0 & 61.4 & 54.2 & 53.0 & 61.1 \\[0.1cm]
Qwen-1.5-7B & 71.3 & 67.3 & 56.9 & 65.7 & 69.1 \\
Qwen-1.5-14B & 73.7 & 67.2 & 69.6 & 62.5 & 69.7 \\
Qwen-1.5-32B & 78.0 & 73.7 & 73.6 & 68.3 & 81.3 \\
Qwen-1.5-110B & 78.4 & 76.2 & 72.2 & 69.9 & 80.3 \\[0.1cm]
Qwen-2-7B & 72.2 & 70.5 & 70.4 & 67.9 & 74.1 \\
Qwen-2-57B & 78.2 & 74.1 & 66.7 & 72.9 & 75.2 \\
Qwen-2-70B & 82.1 \notable & 79.5 \notable & 73.6 & 71.7 & \textbf{86.4} \notable \\[0.1cm]
\hline
\multicolumn{6}{l}{\textbf{01/Yi(零一万物) Series}} \\
Yi-1.5-6B & 66.8 & 62.7 & 59.7 & 63.0 & 63.1 \\
Yi-1.5-9B & 72.4 & 69.2 & 63.9 & 65.1 & 73.5 \\
Yi-1.5-34B & 80.1 & 76.6 & 66.7 & 71.1 & 75.9 \\[0.1cm]
\hline
\multicolumn{6}{l}{\textbf{Deepseek(深度求索) Series}} \\
Deepseek-67B & 69.7 & 69.5 & 61.8 & 62.8 & 64.3 \\[0.1cm]
\hline
\multicolumn{6}{l}{\textbf{Zhipu(智谱)/ChatGLM Series}} \\
ChatGLM3-6B & 55.6 & 53.3 & 47.9 & 53.9 & 54.0 \\
GLM-4-9B & 73.0 & 67.1 & 61.1 & 65.1 & 72.8 \\[0.1cm]
\hline
\multicolumn{6}{l}{\textbf{PJ Lab(浦语)/InternLM(书生) Series}} \\
InternLM-2.5-7B & 71.1 & 63.8 & 61.1 & 61.4 & 64.2 \\[0.1cm]
\hline
\end{tabular}
\end{table*}

\begin{figure*}
\centering
\includegraphics[width=1\textwidth]{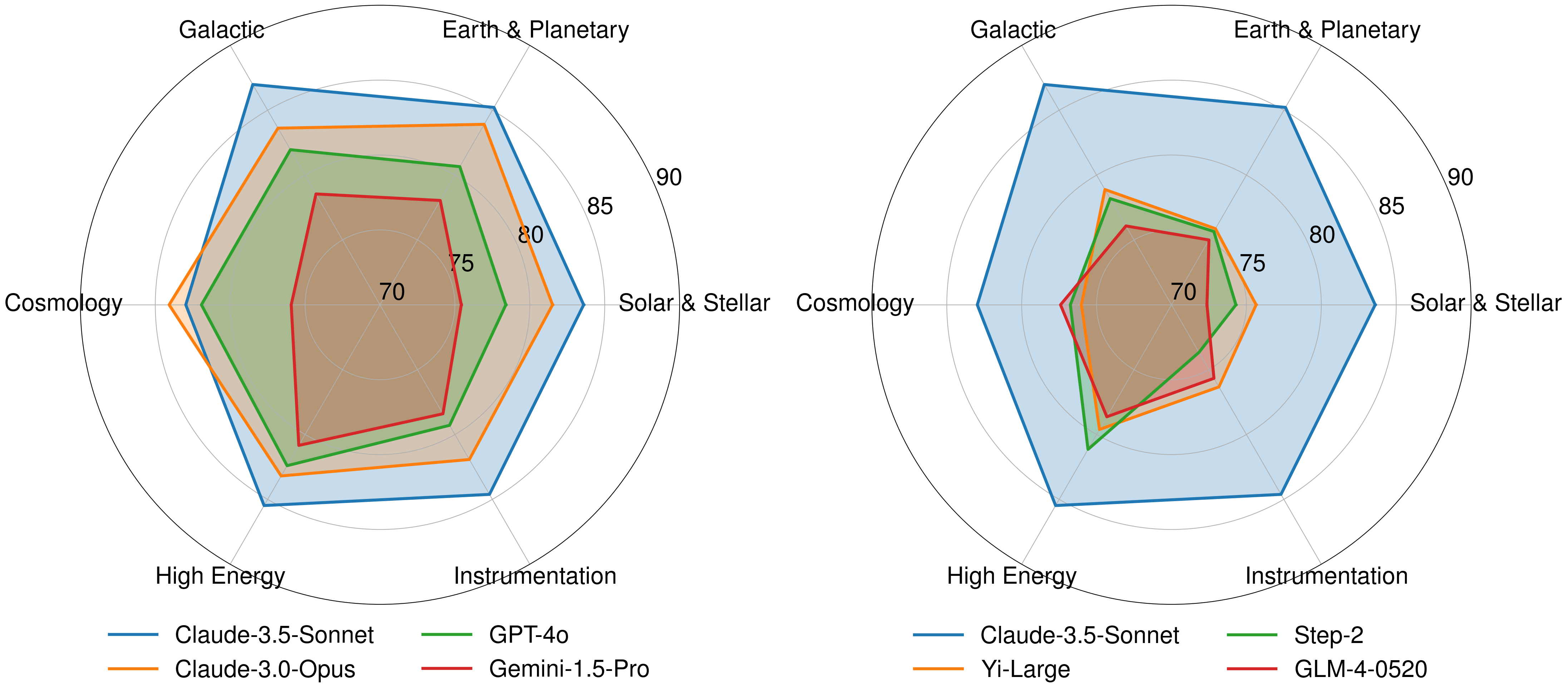}
\caption{Performance of Selected Proprietary Large Language Models on Astronomy Multiple Choice Questions by Subfield Topic. The results are shown in the radar chart, with concentric circles representing different performance levels. The six categories for topics follow the subcategorization of the ArXiv astro-ph classification: `Solar and Stellar Astrophysics', `Earth and Planetary Astrophysics', `Astrophysics of Galaxies', `Cosmology and Nongalactic Astrophysics', `High Energy Astrophysics', and `Instrumentation and Methods for Astrophysics'. The left panel shows the results from Claude-3.5-Sonnet, GPT-4o, Claude-3.0-Opus, and Gemini-1.5-Pro, and the right panel features Yi-Large, Step-2, and GLM-4-0520. Despite these models performing on par with each other in general benchmarks, the latter group seems to perform worse on specialized and somewhat niche astronomical topics. There is a tentative trend of more limitations in recent astronomical research topics such as `Solar and Stellar Astrophysics', `Earth and Planetary Astrophysics', and `Instrumentation and Methods for Astrophysics'. This suggests that part of the degradation might correlate with the training sets adopted in these different models, affecting their performance on more specialized topics. The full results of all other models are listed in Table~\ref{table3}.}
\label{fig4}
\end{figure*}

Fig.~\ref{fig4} illustrates the performance comparison of different proprietary models across various topic classifications, with detailed results shown in Table~\ref{table3}. The medal tally in this table represents the top three performers for each category, with gold, silver, and bronze medals awarded respectively. Blue medals indicate honorable mentions, ranking 4-6 or tied with 6th place. The medals are given considering both the proprietary models and open-weights models (which we will discuss in Section~\ref{sec:open-models}, also see Table~\ref{table4} for the results for open-weights models).

The left panel showcases the performance of some of the strongest proprietary models, including Claude-3.5-Sonnet, GPT-4o, Claude-3.0-Opus, and Gemini-1.5-Pro. As evident from Table~\ref{table3}, Claude-3.5-Sonnet consistently outperforms others, achieving gold medals in five out of six categories and a silver in Cosmology \& Nongalactic Astrophysics (83.0\% compared to Claude-3.0-Opus's 84.1\%). 

As evident from the panel, these high-performing models demonstrate no obvious weaknesses across topics. They tend to perform consistently well in all areas, with a gradual improvement observed from GPT-4o to Claude-3.0-Opus and Claude-3.5-Sonnet. Gemini-1.5-Pro, despite its strong performance in general benchmarks, shows surprising weaknesses in Cosmology \& Nongalactic Astrophysics (75.9\%) and Solar \& Stellar Astrophysics (75.4\%), highlighting the importance of domain-specific evaluations. 

The left panel of Fig~\ref{fig4} also illustrates that weaker models often exhibit greater deficits in Earth and Planetary Astrophysics, Solar and Stellar Astrophysics, and Instrumentation and Methods for Astrophysics. This pattern might suggest a correlation between model degradation and the volume of available training data for each topic. Astrophysics of Galaxies, for example, arguably has the most extensive training data, partly due to the long history of galactic studies and the field's interconnectedness with other domains such as stellar astrophysics and cosmology. Conversely, newer fields like exoplanet studies, or area that rely more on historical context such as stellar astrophysics and instrumentation and methods, appear to suffer the most when transitioning to weaker versions of proprietary models. 

\begin{figure*}
\centering
\includegraphics[width=1\textwidth]{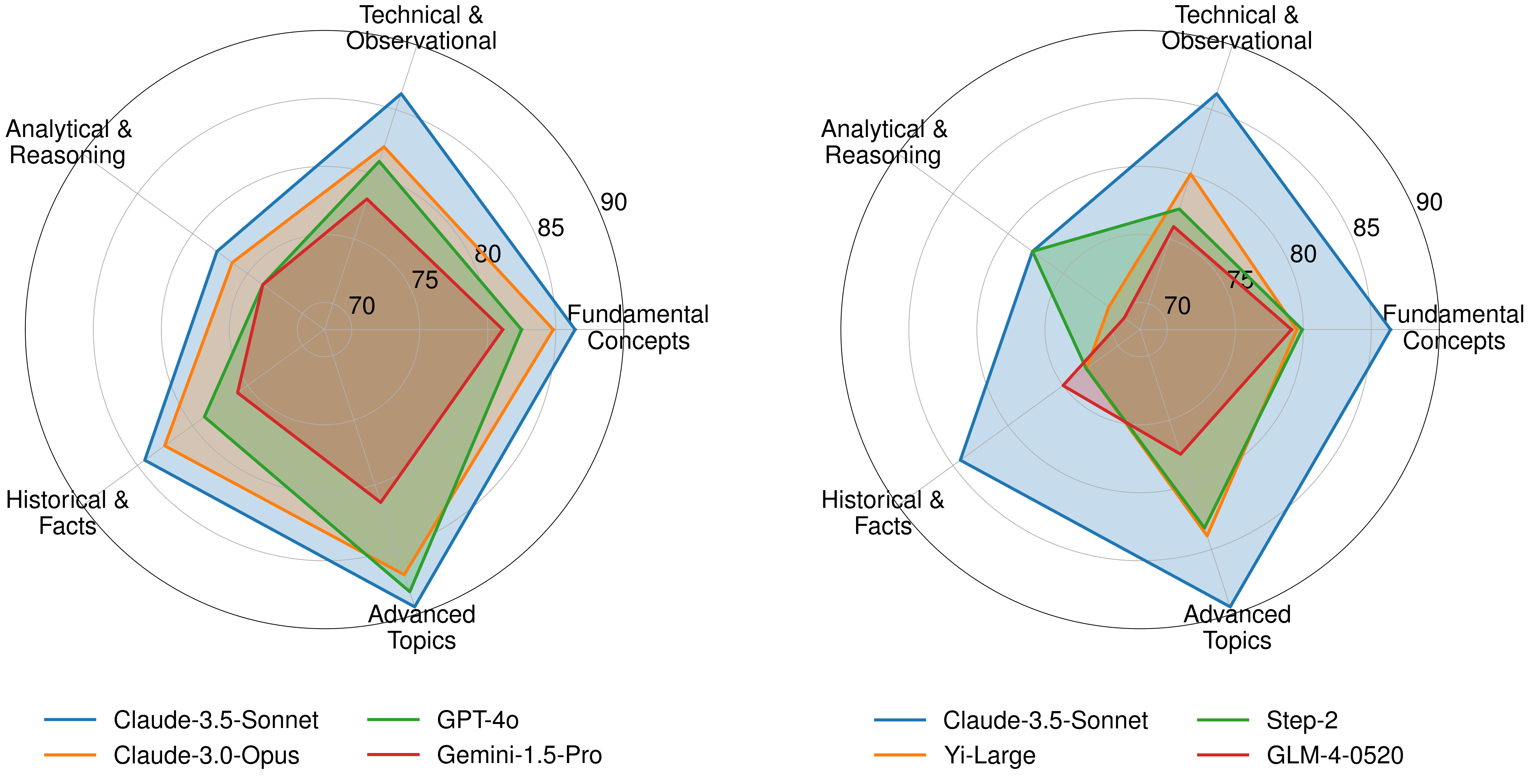}
\caption{Performance of Selected Proprietary Large Language Models by Tested Ability on Astronomy Multiple Choice Questions. This plot is similar to Fig.\ref{fig4}, except here we categorize the questions based on their tested ability rather than subfield topics. The five classes are `Understanding Fundamental Concepts', `Technical and Observational Techniques', `Analytical and Reasoning Skills', `Historical and Theoretical Knowledge', and `Current Research and Advanced Topics'. Despite these models performing comparably in general benchmarks, as shown in the right panel, the non-English-focused models exhibit more significant degradation in `Historical and Theoretical Knowledge', and occasionally in `Current Research and Advanced Topics' compared to the English-focused models as shown on the left panel. This further suggests that the limitations observed in Fig.\ref{fig4} may stem from differences in training data among these models, affecting their performance on more specialized astronomical topics. The complete results for all other models are presented in Table~\ref{table5}.}
\label{fig5}
\end{figure*}

The right panels of Fig.~\ref{fig4} and \ref{fig5} (alo see Table~\ref{table3}) present the performance of top-tier models from non-primarily-English-based models, including Yi-Large, Step-2, and GLM-4-0520. Despite these models performing comparably in general benchmarks, they show weaker performance on specialized astronomical topics. For instance, Yi-Large, the strongest among them with an average score of 77.3\%, achieves scores ranging from 75.6\% to 79.6\% across categories, notably lower than Claude-3.5-Sonnet's range of 83.0\% to 87.0\%. Step-2\footnote{As we tested this model in early July 2024, Step-2 is still under ``nightly" version, and might continue to update and improve.} is a close second among proprietary non-English-focused models. With an average score of 76.6\%, it achieves ``notable mention" in two topics, but falls short compared to Yi-Large due to weaknesses in other topics. Further, Fig.~\ref{fig4} reveals a tentative trend of more pronounced limitations in astronomical research topics aforementioned. For example, GLM-4-0520 scores 72.4\% in Solar \& Stellar Astrophysics, 75.0\% in Earth \& Planetary Astrophysics and 75.7\% in Instrumentation and Methods for Astrophysics, weaker than the other three topics. 

Apart from separating into subfield topics, Fig.~\ref{fig5}, complemented by Table~\ref{table5}, provides insights into different abilities tested for proprietary models. While the left panel shows that weaker models in the English-focused series generally degrade homogeneously across categories, Non-English-focused models on the right on the other hand, demonstrate a marked degradation in `Historical \& Theoretical Knowledge'. For instance, GLM-4-0520 scores 75.0\% in this category, compared to Claude-3.5-Sonnet's 84.3\%, most notable among all categories. This trend is also observed in other models, including Yi-Large (72.9\%) and Step-2 (72.9\%). At times, the non-English-focused models also struggle in ``Current Research and Advanced Topics". Together, this might suggest that the limitations stem from differences in training data adopted by developers from different regions, especially lacking data that might either be historical context of astronomy (see related question examples in Appendix~\ref{appendixD}). These differences in data sources arguably might differ  between communities across continents, affecting their performance on more specialized astronomical topics.

Regardless of the model origin, analytical skills remain a notable weakness for most models. In this context, we're referring to logical deductions from observations, rather than complex mathematical problems, as the MCQs are derived from Annual Review of Astronomy and Astrophysics articles. Even the strongest proprietary model, Claude-3.5-Sonnet, achieves only 77.8\% accuracy in Analytical \& Reasoning Skills, compared to its outstanding performance of 86.4\% in Understanding Fundamental Concepts and 89.4\% in Advanced Topics. This gap highlights the ongoing challenges in deploying LLMs as robust astronomical agents, particularly in tasks requiring quick estimations and logical reasoning. 

Interestingly, Step-1 and 2 shows some of the strongest performances in Analytical \& Reasoning Skills, achieving among the highest scores of $\sim 78-79\%$ in this category. Step-2 is a notable trillion-parameter model developed by a team led by experts formerly associated with Microsoft Research Asia. This outlier performance warrants further investigation and could provide valuable insights for improving other models in this critical area. 

The medal tally in Tables~\ref{table3} and \ref{table5} further emphasizes the dominance of Claude-3.5-Sonnet (9 gold, 2 silver). Claude-3.0-Opus (1 gold, 7 silver, 2 bronze, 1 notable mentions) and GPT-4o (1 silver, 2 bronze, 7 notable mentions) are also show strong performance. Several other proprietary models also demonstrate strong performance, often appearing among the top scorers and medalists. Notable models include Yi-Large (2 notable mentions), Step-1 (1 gold) and Step-2 (1 bronze, 2 notable mentions), Gemini-1.5-Pro (2 notable mentions) as well as Claude-3.0-Sonnet (2 notable mentions) and Claude-3.0-Haiku (4 notable mentions), showcasing the diversity of strengths among different AI developers in the field of astronomical knowledge.

\begin{figure*}
\centering
\includegraphics[width=0.95\textwidth]{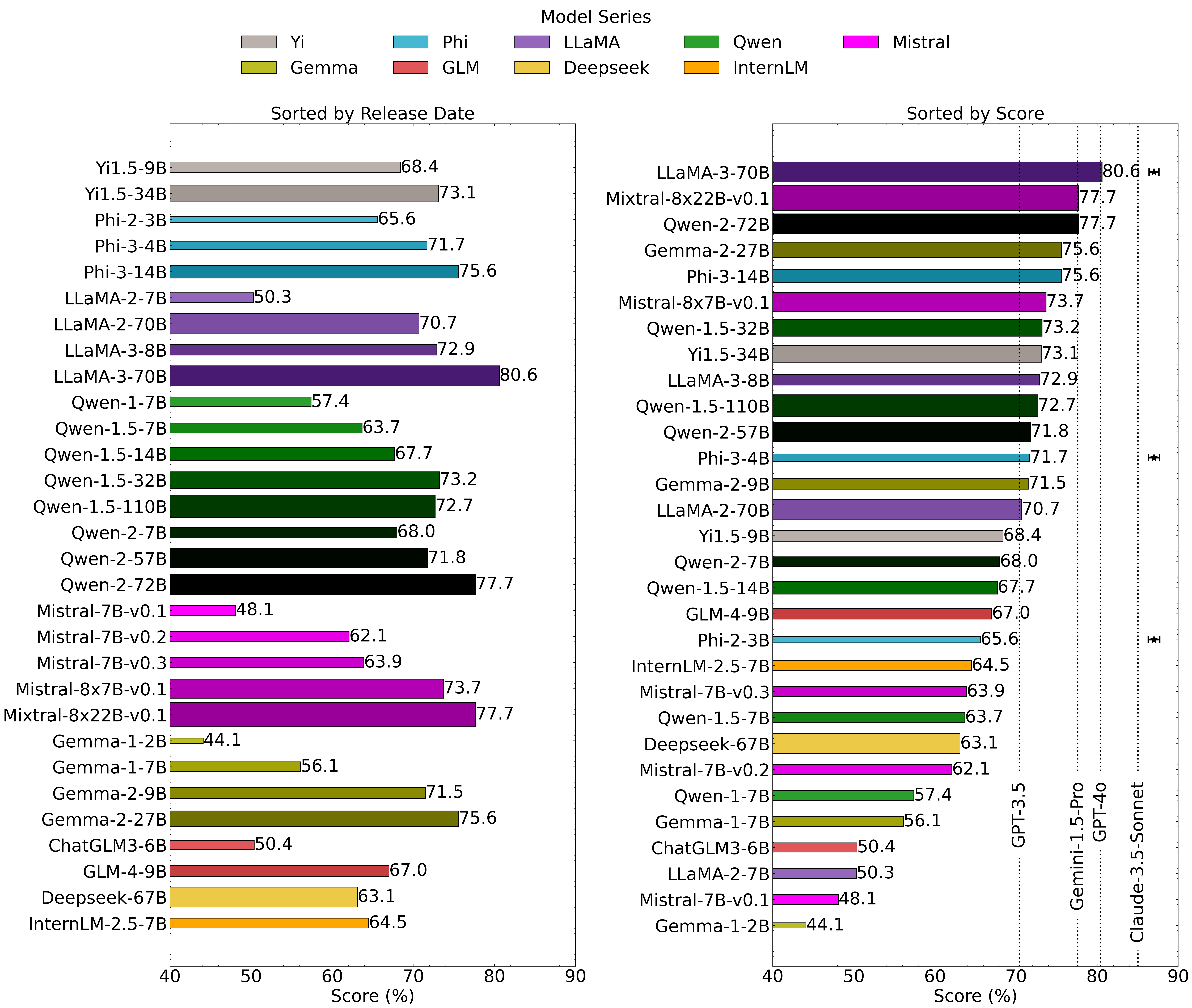}
\caption{Benchmarking scores of open-weights large language models for MCQ answering in astronomical research. The layout is similar to Fig.~\ref{fig1}. We tested Yi-1.5 (9B, 34B), Phi (2-3B, 3-4B, 3-14B), LLaMA (2-7B, 2-70B, 3-8B, 3-70B), Qwen (1-7B, 1.5-7B, 1.5-14B, 1.5-32B, 1.5-110B, 2-7B, 2-57B, 2-72B), Mistral (7B-v0.1, 7B-v0.2, 7B-v0.3), Mixtral (8x7B-v0.1, 8x22B-v0.1), Gemma (1-2B, 1-7B, 2-9B, 2-27B), GLM (ChatGLM3-6B, 4-9B), Deepseek-67B and InternLM-2.5-7B. The left panel shows the score sorted by model series. The right panel shows the same scores sorted by overall performance, regardless of model series. Also in the right panel, the scores for four reference proprietary models (Claude-3.5-Sonnet, GPT-4o, Gemini-1.5-Pro, and GPT-3.5) are plotted as vertical reference lines. The size of the bar is scaled with the number of parameters of the models. LLaMA-3-70B performs best with an 80.6\% accuracy, outperforming even GPT-4o (80.4\%) and Gemini-1.5-Pro (77.6\%), although it is still worse than Claude-3.5-Sonnet (85.0\%). Qwen-2-72B and Mixtral-8x22B-v0.1 are also competitive with 77.7\% accuracy, which is on par with Gemini-1.5-Pro. This shows that open-weights models, at least those with 70 billion parameters or more, are competitive in terms of astronomical Q\&A performance. The error bars in the right panel display the statistical uncertainties for three representative models.}
\label{fig6}
\end{figure*}

\begin{figure*}
\centering
\includegraphics[width=1\textwidth]{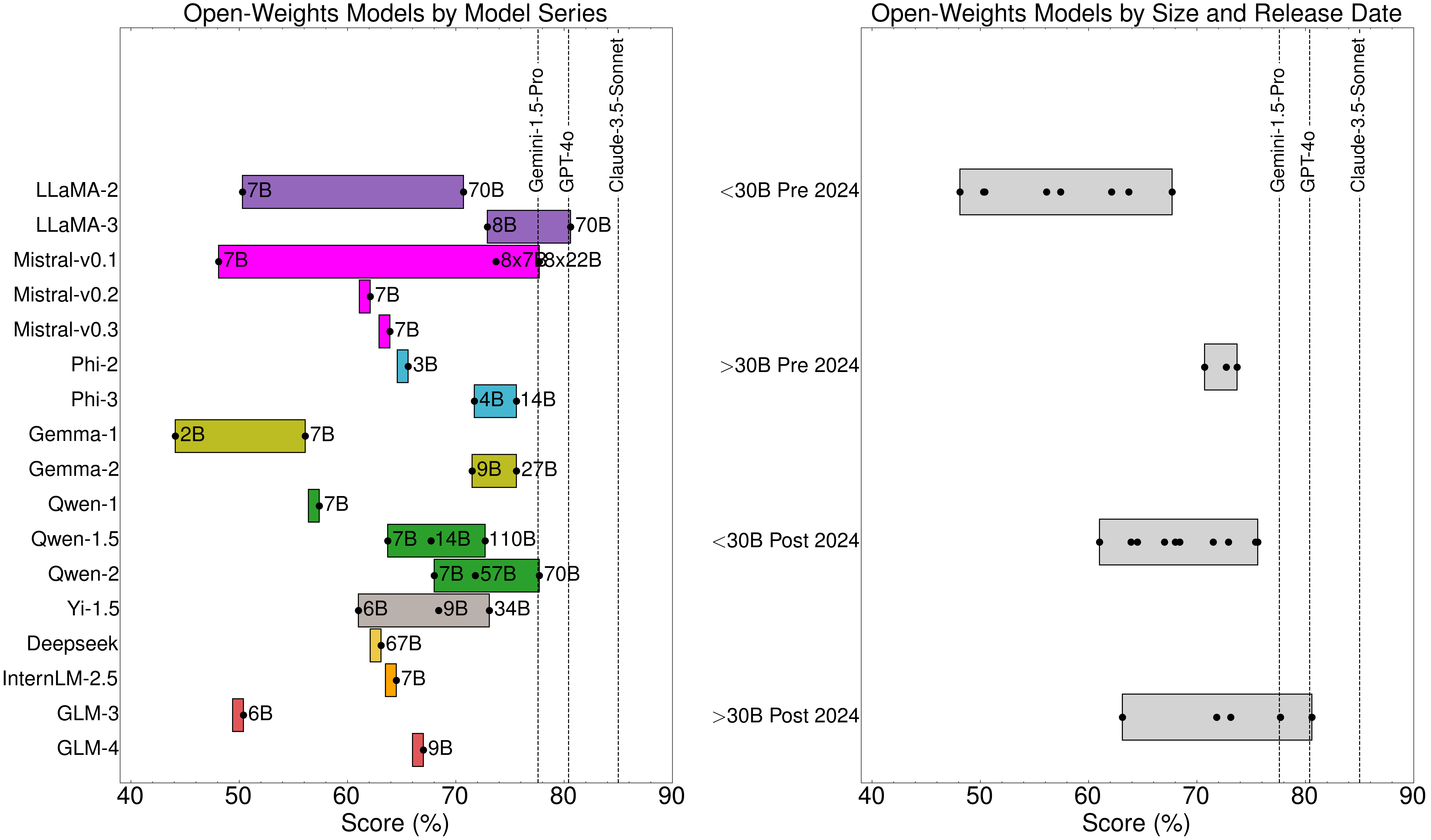}
\caption{The evaluation of open-weights models in astronomical answering benchmarking compared to proprietary benchmarks: Gemini-1.5-Pro (77.6\%), GPT-4o (80.4\%), and the state-of-the-art Claude-3.5-Sonnet (85.0\%). The left panel displays different model series, including LLaMA, Mistral, Phi, Gemma, Qwen, Yi, GLM, InternLM and Deepseek, adopting the same color coding as in Fig.~\ref{fig6}. The various model series show marked differences over a short succession of time, with lightweight models (e.g., LLaMA-3-8B at 72.9\%) often surpassing models 10 times larger from previous series (e.g., LLaMA-2-70B at 70.7\%). Notably, the largest models from Meta's LLaMA-3-70B (80.6\%), Alibaba's Qwen-2-72B (77.7\%), and Mistral (Mixtral 8x22B-v0.1 at 77.7\%) perform on par with or better than Gemini-1.5-Pro and GPT-4o, although still falling behind the state-of-the-art Claude-3.5-Sonnet. The right panel illustrates the evolution of models at different release times (pre and post-2024) and sizes ($<$30B to $>$30B). Note that release times are approximate (see text for details). The overall trend is clear: the best-performing lightweight ($<$30B) models in 2024 have already surpassed the $>30$B-scale models from before 2024, and the best performing heavyweight ($>$30B) models in 2024 are performing on par with some of the existing proprietary models.}
\label{fig7}
\end{figure*}

\section{Benchmarking Open-Weights Large Language Models}
\label{sec:open-models}

While proprietary models have been advancing at a remarkable pace, relying on API calls to perform large-scale astronomical research may remain cost-prohibitive or challenging to justify to survey management. open-weights models are critical for LLM deployment in astronomical research for two primary reasons: (1) In academic settings, it is often easier to secure GPU compute resources than to obtain grants, especially as the deployment of proprietary LLM APIs is not yet widely adopted in astronomical research. (2) open-weights models offer the possibility of further continual pretraining or specialized fine-tuning, potentially allowing for better optimization for downstream tasks specific to astronomy. The latter point is a key motivation for AstroMLab\footnote{astromlab.org}, which constitutes the core developers for AstroLLaMA \citep{Nguyen2023} and AstroLLaMA-chat \citep{Perkowski2024}. We will defer the detailed benchmarking of these astro-series large language models to forthcoming papers.

\subsection{Comparative Analysis of Leading Open-Weights Model Series}

In this section, we present a comprehensive evaluation of prominent open-weights models, ranging from smaller-scale 2B parameter models to larger 176B parameter models. Our analysis aims to trace the evolution of these open-weights models from our astronomical perspective. We focus on several representative model series:

\begin{enumerate}
\item Microsoft's Phi-2 (3B) and Phi-3 (4B, 14B) models \citep{Phi2023,Phi2024}
\item Meta's LLaMA-2 (7B, 70B) and LLaMA-3 (8B, 70B) \citep{Llama2023a,Llama2023B}
\item Alibaba's Qwen-1 (7B), Qwen-1.5 (7B, 14B, 32B, 110B), and Qwen-2 (7B, 57B, 72B) \citep{Qwen2023}
\item MistralAI's Mistral (7B-v0.1, 7B-v0.2, 7B-v0.3) and Mixtral MOE (8x7B-v0.1, 8x22B-v0.1) models \citep{Mistral2023,Mistral2024}
\item The Yi-1.5 series (6B, 9B, 34B) \citep{01AI2024}
\item Google's Gemma-1 (2B, 7B) and Gemma-2 (9B, 27B) series \citep{Gemma2024}
\item Zhipu AI's ChatGLM3 (6B) and GLM-4 (9B) series \citep{GLM2024}
\item DeepSeek AI's 67B model \citep{Deepseek2024}
\item PJ Lab's InternLM-2.5 model (7B) \citep{Internlm2024}
\end{enumerate}

This selection encompasses some of the most recognized open-weights models from myriad of developers. It provides a diverse representation of global efforts in open-weights AI development. Additionally, some models, including Yi, GLM and DeepSeek, adopt a hybrid approach that combines proprietary and open-weights elements. To provide a comprehensive comparison, we have evaluated the open-weights versions and proprietary versions of these models separately.

Fig.~\ref{fig6} presents a comprehensive score comparison in histogram form, similar to Fig.~\ref{fig1}. The left panel groups models by series, with darker shades indicating more recent or larger models within each series. The width of individual bar is scaled (logarithmically) with the number of parameters in these open-weights models. This arrangement allows for easy comparison of performance evolution within each model series. The right panel shows the same scores sorted by overall performance, regardless of model series, providing a clear view of how different models stack up against each other across all model series. While our primary focus is on open-weights models, we have included several representative proprietary models for reference and comparison. These proprietary models—Claude-3.5-Sonnet, GPT-4o, Gemini-1.5-Pro, and GPT-3.5—are represented with vertical reference lines.

The detailed performance of all these models and their score differences compared to the four proprietary benchmarks are also shown in Table~\ref{table2}. To better understand the evolution of model performance, we plot that in Fig.~\ref{fig7}, which illustrates the progression of open-weights models compared to proprietary benchmarks. The left panel, which displays the performance of individual model series as separate bar charts, reveals interesting trends across different model series. 

The LLaMA series shows consistent improvement from version 2 to 3, with even the 8B model of LLaMA-3 (72.9\%) outperforming the 70B model of LLaMA-2 (70.7\%). The Mistral series demonstrates the power of mixture-of-experts (MOE) architecture, with Mixtral-8x22B-v0.1 (77.7\%) performing significantly better than the standard Mistral models, such as Mistral-7B-v0.1 (48.1\%), Mistral-7B-v0.2 (62.1\%) and Mistral-7B-v0.3 (63.9\%). The Qwen series shows a general trend of improvement across versions and sizes, with Qwen-2-70B achieving top-tier performance (77.7\%). Similar improvement trends are observed in other series, notably from Phi-2-3B (65.6\%) to Phi-3-14B (75.6\%), and Gemma-1-7B (56.1\%) to Gemma-2-27B (75.6\%). 

The right panel of Fig.~\ref{fig7} provides an overall view of model evolution across different release times (pre and post-2024) and sizes ($<$30B to $>$30B).  We note that, this categorization is not strictly defined by the calendar year. Some models released in early 2024 but superseded by June 2024 (when this paper was written) are categorized as pre-2024 for the purposes of this analysis. In particular, the pre-2024 $<$30B models include LLaMA-2-7B, Mistral-7B-v0.1, Mistral-7B-v0.2, Qwen-1-7B, ChatGLM3-6B, Gemma-1-7B, Qwen-1.5-7B, and Qwen-1.5-14B. The pre-2024 $>$30B models include LLaMA-2-70B, Mistral-8x7B-v0.1, and Qwen-1.5-110B. The post-2024 $<$30B models include LLaMA-3-8B, Mistral-7B-v0.3, Phi-3-14B, Gemma-2-9B, Gemma-2-27B, Qwen-2-7B, Yi-1.5-6B, Yi-1.5-9B, GLM-4-9B, and InternLM-2.5-7B. The post-2024 $>$30B models include LLaMA-3-70B, Mistral-8x22B-v0.1, Qwen-2-57B, Qwen-2-70B, Yi-1.5-34B, and Deepseek-67B.

Pre-2024 models like LLaMA-2-70B (70.7\%) and Mistral-8x7B-v0.1 (73.7\%) show performance comparable to GPT-3.5 (70.4\%). However, post-2024 models, particularly those in the $>$70B parameter range, show marked improvements. LLaMA-3-70B (80.6\%), Qwen-2-70B (77.7\%), and Mixtral 8x22B-v0.1 (77.7\%) perform on par with or better than Gemini-1.5-Pro (77.6\%) and even GPT-4o (80.4\%) in the case of LLaMA-3-70B. This is a significant development for the astronomical community, as it suggests that researchers can leverage these open-weights models to achieve performance comparable to some of the best proprietary options, potentially at a fraction of the cost and with greater flexibility for customization and fine-tuning.

The rapid progress in open-weights model development is evident in the drastic performance improvements over short periods. For example, Mistral-7B-v0.1 scored only 48.1\%, while its successor Mistral-7B-v0.3 achieved 63.9\%, a 15.8 percentage point improvement. Similarly, Qwen-1-7B scored 57.4\%, while Qwen-2-7B reached 68.0\%, showing a 10.6 percentage point increase.

As models with $>$30B parameters are often prohibitively costly in academic settings, it's crucial to understand the performance of models with $<$30B parameters. Unsurprisingly, these smaller models perform worse than their larger counterparts. The best $<$10B parameter model, LLaMA-3-8B, achieves an accuracy of 72.9\%, still far below the top proprietary models. In the $10-30$B parameter range, however, Gemma-2-27B and Phi-3-14B both achieve accuracies of 75.6\%, closing the gap to the lower end of the proprietary models like Gemini-1.5-Pro (77.6\%).

As we noted, some non-English-focused models while performing well in other benchmarks, seem to fare less favorably in this astronomy benchmark, especially in the smaller-scale competition. For instance, Yi-1.5-34B achieves 73.1\%, Deepseek-67B reaches 63.1\%, and GLM-4-9B attains 67.0\%. Similarly, while Qwen-2-72B is one of the stronger open-weights models, only about 2.9 points behind LLaMA-3-70B at the 70 billion parameters level, Qwen-2-7B only reaches 68.0\%, all lagging behind LLaMA-3-8B's 72.9\%. Nonetheless, the improvement within series is notable, as seen in the comparison between ChatGLM3-6B (50.4\%) and GLM-4-9B (67.0\%), showing a 16.6 percentage point increase. 

We note that proprietary models often undergo more rigorous customization and testing, which may contribute to their performance edge. Claude-3.5-Sonnet still maintains a lead of 4.4 percentage points over the closest open-weights counterpart (LLaMA-3-70B). However, the fact that LLaMA-3-70B achieves parity with and slightly surpasses GPT-4o in our benchmark is a significant achievement. In summary, while open-weights models, particularly at the $>$70 billion parameter scale, are approaching the performance of leading proprietary models, significant work remains to ensure more affordable models suitable for academic settings can be competitive as LLM agents in astronomical research.
\begin{figure*}
\centering
\includegraphics[width=1\textwidth]{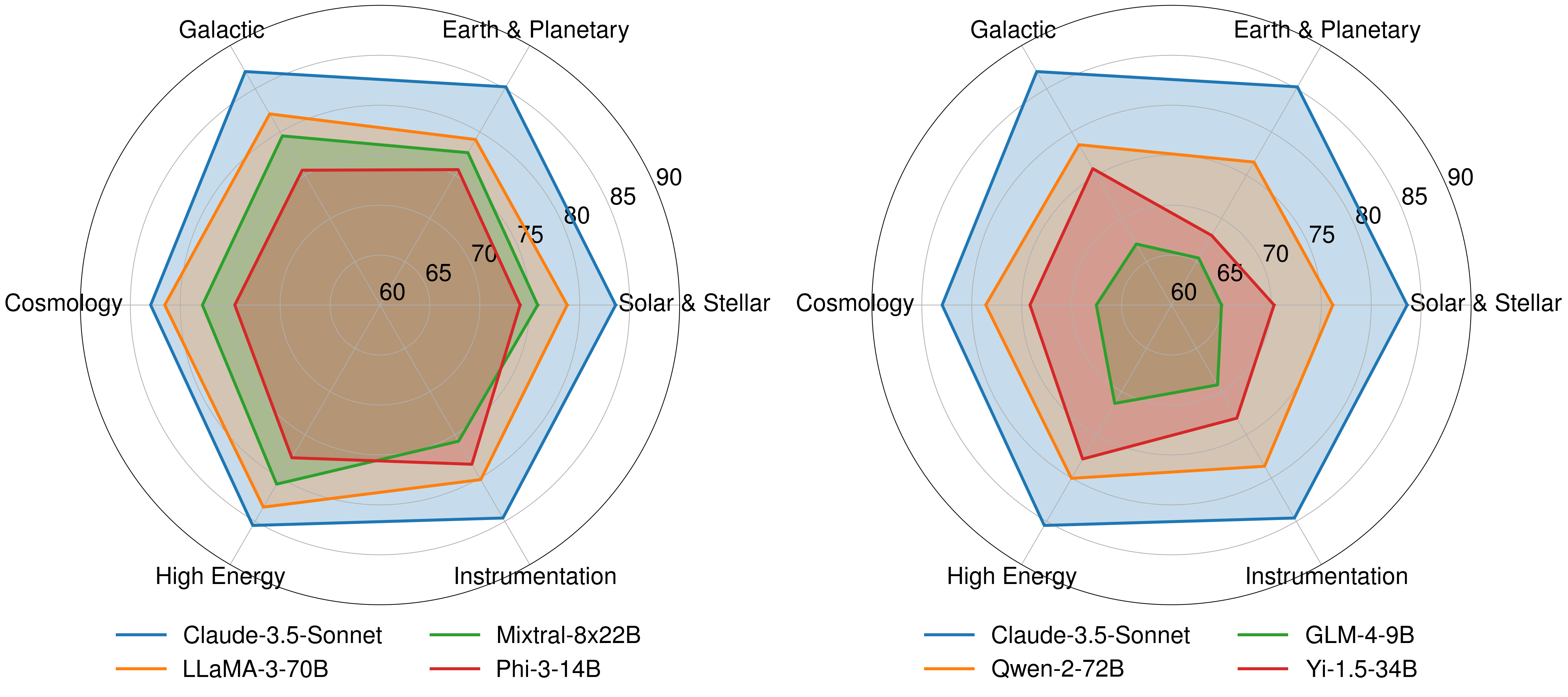}
\caption{Performance of Selected Open-Weights Large Language Models on Astronomy Multiple Choice Questions by Subfield Topic. This plot is similar to Fig.\ref{fig5}, but focuses on open-weights models instead of proprietary models. The left panel shows results from LLaMA-3-70B, Mixtral-8x22B-v0.1, and Phi-3-14B, while the right panel features Qwen-2-72B, Yi-1.5-34B, and GLM-4-9B. Both panels include Claude-3.5-Sonnet as a reference for the best-performing model. Open-weight models exhibit notable performance degradation in more recent topics related to 'Solar and Stellar Astrophysics', 'Earth and Planetary Astrophysics', and 'Instrumentation and Methods for Astrophysics'. This trend is visible in English-focused models (left panel) and becomes more pronounced in non-English-focused models (right panel). The complete results for all other models are presented in Table\ref{table4}.}
\label{fig8}
\end{figure*}

\subsection{Analyzing Performance Discrepancies in Open-Weights Models}

\begin{figure*}
\centering
\includegraphics[width=1\textwidth]{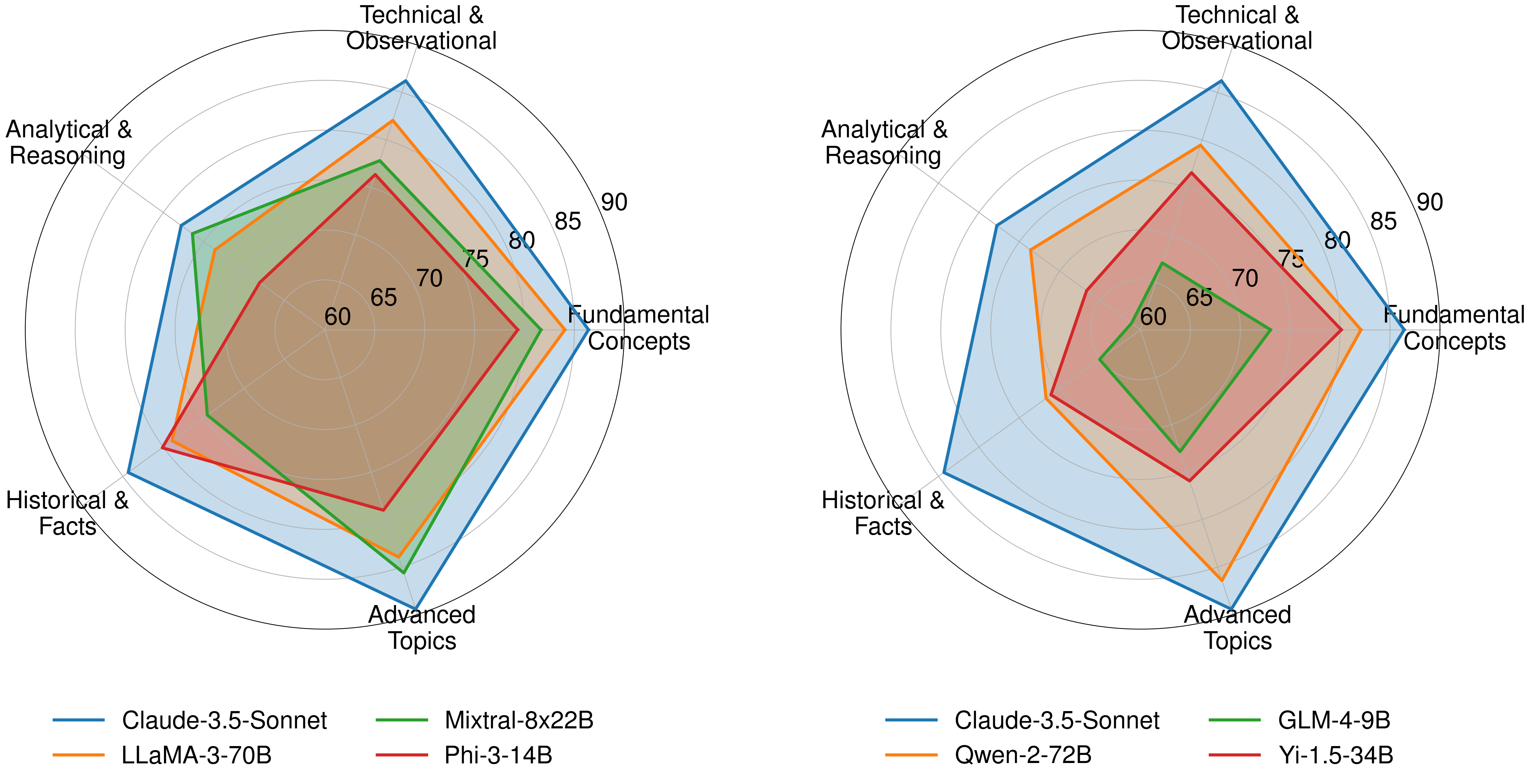}
\caption{Performance of Selected Open-Weights Large Language Models by Tested Ability on Astronomy Multiple Choice Questions.
The plot is similar to Fig.~\ref{fig5} except here we study the open-weights models instead of the proprietary models. The left panel shows LLaMA-3-70B, Mixtral-8x22B-v0.1, and Phi-3-14B, and the right panel features Qwen-2-72B, Yi-1.5-34B, and GLM-4-9B. For both panels, the best-performing model, Claude-3.5-Sonnet, is shown as a reference. Similar to Fig.~\ref{fig5}, there is a notable degradation in performance for the ability related to `Historical and Theoretical Knowledge', and occationally `Current Research and Advanced Topics', compared to the state-of-the-art Claude-3.5-Sonnet. This is more pronounced for the non-English-focused models (right panel) but is also visible in English-focused models (left panel). This further confirms that the limitations might stem from the training data adopted in these different models. The full results of all other models are listed in Table~\ref{table6}.}
\label{fig9}
\end{figure*}

Following the classification scheme described in Fig.\ref{fig4} and \ref{fig5}, we assess why some open-weights models perform weaker than others. Our analysis is divided into two parts, illustrated in Fig.~\ref{fig8} and \ref{fig9}. Fig.\ref{fig8} focuses on the accuracy across different subfield topics, while Fig.\ref{fig9} examines the abilities tested, mirroring the approach in Fig.\ref{fig4} and \ref{fig5}. For reproducibility, all detailed accuracies for each question class are summarized in Tables~\ref{table5} and \ref{table6}.

The left panels of Fig.~\ref{fig8} and \ref{fig9} present a comparison between the best-performing proprietary model (Claude-3.5-Sonnet) and top-performing open-weights models (LLaMA-3-70B, Mixtral-8x22B-v0.1, and Phi-3-14B). Claude-3.5-Sonnet demonstrates consistently high performance across all topics and abilities, with scores ranging from 77.8\% to 89.4\%. In contrast, even the best open-weights models show some variability, particularly in more recent or rapidly evolving fields. Notably, these open-weights models generally perform weaker in Earth and Planetary Astrophysics, Solar and Stellar Astrophysics, and Instrumentation and Methods for Astrophysics. 

This pattern echoes our previous observations and suggests that these topics, being more recent (exoplanets) or requiring more historical context (stellar astrophysics, instrumentations and methods), might have less representation in the training data. For instance: (1) LLaMA-3-70B shows lower scores in Earth and Planetary Astrophysics (79.1\%) and Solar and Stellar Astrophysics (78.7\%) compared to its performance in other areas. (2) Mixtral-8x22B-v0.1, demonstrates relatively weaker performance in all these three areas with scores 75.7\% to 77.6\% compared to the other topics. (3) Phi-3-14B shows the weakest performance in Earth and Planetary Astrophysics (74.0\%). (4) Gemma-2-27B also shows the weakest performance in Earth and Planetary Astrophysics (73.3\%) and Instrumentations and Methods (73.4\%).

The left panel of Fig.~\ref{fig9} further reinforces this trend when examining the abilities tested. Open-weights models generally show more decrements in ``Historical and Theoretical Knowledge" and ``Current Research and Advanced Topics" compared to Claude-3.5-Sonnet. This discrepancy is particularly evident in Phi-3-14B on ``Current Research and Advanced Topics" (hence the weak performance" and Mixtral-8x22B-v0.1 on ``Current Research and Advanced Topics". For instance, Phi-3-14B scores 79.0\% in Advanced Topics, while Claude-3.5-Sonnet achieves 89.4\% respectively, showing a gap of 10.4 percentage points. Similarly, Mixtral-8x22B-v0.1 scores 74.6\% in Historical and Theoretical Knowledge, demonstrating a substantial 9.7 percentage point lag in historical knowledge compared to Claude-3.5-Sonnet's 84.3\%. Even LLaMA-3-70B, while performing better, still shows noticeable deficits with about 5 percentage points in both of these categories.

The limitations potentially due to training data are even more pronounced for some state-of-the-art models that have been trained on datasets not primarily focused on English-language content. The right panels of Fig.~\ref{fig8} and \ref{fig9} feature top-performing open-weights models from various developers: Qwen-2-70B, Yi-1.5-34B, and GLM-4-9B. These models exhibit notable weaknesses in the aforementioned areas. In Fig.~\ref{fig8}, we observe that these models struggle particularly with Earth and Planetary Astrophysics, Solar and Stellar Astrophysics, and Instrumentation and Methods for Astrophysics. For instance: (1) Qwen-2-70B, while strong overall, shows a dip in Earth and Planetary Astrophysics (76.5\%) and Solar and Stellar Astrophysics (76.1\%) compared to its performance in other areas.(2) Yi-1.5-34B demonstrates lower scores in Earth and Planetary Astrophysics (68.0\%) and Solar and Stellar Astrophysics (70.3\%). (3) GLM-4-9B, the smallest of the three, shows consistent weaknesses across these areas, with also notably low scores in Solar and Stellar Astrophysics (65.0\%) and Earth and Planetary Astrophysics (65.4\%).

Fig.~\ref{fig9} further highlights these models' limitations in Historical and Theoretical Knowledge and Current Research and Advanced Topics: (1) Qwen-2-70B shows a visible weakness in Historical \& Theoretical Knowledge (71.7\%), a gap of 12.6 percentage points compared to Claude-3.5-Sonnet (84.3\%). (2) Yi-1.5-34B also falls short in Historical \& Theoretical Knowledge (71.1\%) and Current Research and Advanced Topics (75.9\%), showing gaps of 13.2 and 13.5 percentage points respectively compared to Claude-3.5-Sonnet. (3) GLM-4-9B lags behind also, particularly in Historical \& Theoretical Knowledge (65.1\%) and Advanced Topics (72.8\%), with gaps of 19.2 and 16.6 percentage points respectively compared to Claude-3.5-Sonnet.

These patterns confirm what we discussed in the previous section, suggesting that the challenges in acquiring and effectively utilizing comprehensive training data for rapidly evolving or historically rich astronomical topics are universal across different development regions. It's worth noting that although these models also perform weaker in Analytical \& Reasoning Skills, this seems to be a generic trend even for Claude-3.5-Sonnet (77.8\%). The gap in analytical skills is less pronounced compared to other categories, suggesting this might be a common challenge across both open-weights and proprietary models.

Finally, the medal tally in Tables~\ref{table4} and \ref{table6} highlights the strong performance of several open-weights models. LLaMA-3-70B stands out with 2 silver, 4 bronze, and 4 notable mentions across various categories. Qwen-2-70B demonstrates competitive performance with 6 notable mentions. Mixtral-8x22B-v0.1 follows with 4 notable mentions. Phi-3-14B secures 1 bronze and 1 notable mention, while Gemma-2-27B earns 2 notable mentions. This distribution of medals and mentions across different model series showcases the advancing capabilities of open-weights models in the field of astronomical knowledge, with some approaching the performance of top proprietary models in certain areas.

\section{Discussion}

This study presents a comprehensive benchmark of proprietary and open-weights LLMs against a high-quality MCQ dataset extracted from the Annual Review of Astronomy and Astrophysics. As one of the first benchmarks specifically assessing LLM capabilities in astronomical research, we provide a thorough baseline, evaluating both current and earlier versions of various proprietary as well as open-weights models to offer insights into the evolution of AI capabilities in this specialized domain.

\subsection{Limitation and rationale behind this benchmarking dataset}

A key question arising from this benchmarking is whether it meaningfully assesses the potential for deploying these models as LLM agents in astronomical research. We emphasize that our Annual Review-driven MCQ dataset is primarily designed to test nuanced understanding within the specialized astronomical research community. This focus is significant given the relatively small footprint of astronomical literature in typical LLM training sets. For context, the entire arXiv astro-ph article corpus comprises less than 3B words, which, even when expanded to include older papers from the Astrophysics Data System, represents only about 0.025 percent of the training set for models like LLaMA-3. While other astronomy-related texts exist (e.g., press releases), the majority lack the detailed knowledge required to answer these MCQ questions accurately.

The closest analog to our MCQ assessment might be the general exams in a graduate programs in astrophysics. In these exams, PhD candidates, after two years of study and coursework, are tested on their broad knowledge of astronomical research to determine their readiness for independent doctoral research. Our MCQ benchmark serves a similar purpose, gauging a model's grasp of foundational and current astronomical knowledge. While this metric provides a valuable assessment of a model's astronomical knowledge base, it represents a necessary but not sufficient condition for evaluating a model's capacity to conduct actual astronomical research tasks. 

Nonetheless, we emphasize that the benchmark presented in this study is representative of the current deployment of LLM agents in astronomical research. Tasks such as knowledge graph generation \citep{Sun2024} or policy learning through LLM agents are critically and predominantly determined by the model's ability to perform exact astronomical knowledge recall, which largely inspired this study. This benchmark, therefore, offers a crucial baseline for understanding how well individual models comprehend and can apply astronomical concepts. However, it should be considered alongside other metrics when assessing a model's full potential for conducting or assisting in astronomical research.

Our new benchmark represents a significant addition to current LLM evaluation efforts. This is evident in the stark contrast between our results and those of other well-known benchmarks discussed in the introduction. While various top contending models such as GPT-4o, Claude-3.5-Sonnet, GLM-4-0520 and Gemini-1.5-Pro can be comprable in tasks across various general benchmarks, our evaluation reveals surprising inefficiencies in models like GPT-4o, Gemini-1.5-Pro and GLM-4-0520, which can vary by three orders of magnitude value-efficient for this specific astronomical benchmark.

This discrepancy is not entirely unexpected, given the minority status of astrophysics in vast training datasets. The specialized knowledge required for our benchmark is likely more brittle and heavily dependent on the specific training data used, which may explain the observed performance variations. This finding underscores the need for more scientific Q\&A benchmarks to better understand LLM agent performance in detailed knowledge recall across various scientific domains. Our pipeline offers an end-to-end approach that could be extended to other Annual Reviews in other scientific fields, and we encourage collaboration in this endeavor.

Future work could explore correlations between performance on this benchmark and a model's ability to engage in more complex, open-ended astronomical research tasks. However, curating such end-to-end astronomical research questions is significantly challenging to automate and would require extensive community input. Additionally, future benchmarks could potentially include more rigorous tests of analytical and reasoning skills specific to astronomical research contexts.

\begin{figure*}
\centering
\includegraphics[width=1\textwidth]{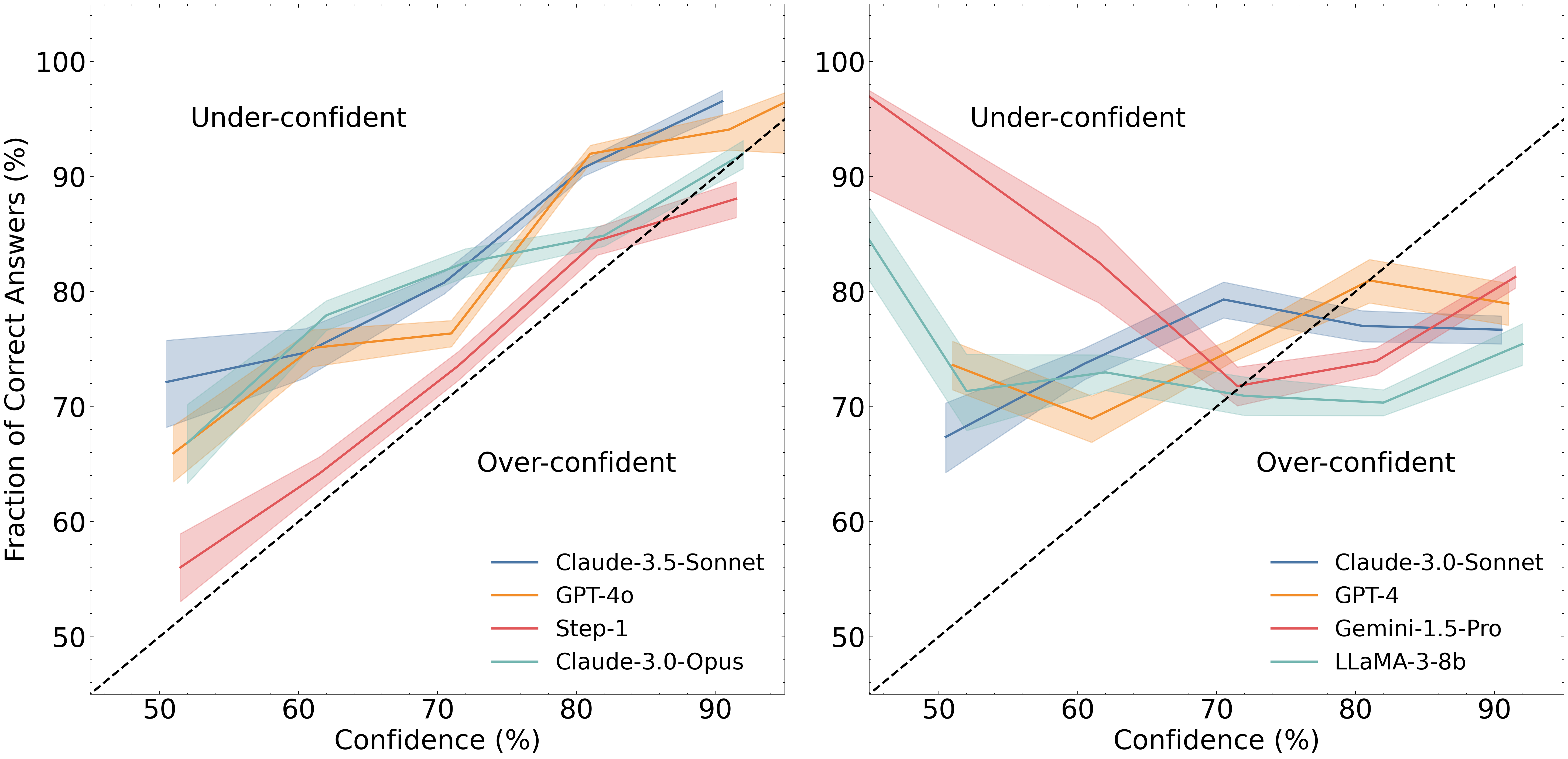}
\caption{Calibration of Confidence in Responses from Large Language Models. We binned the confidence with a bin size of 0.1 from 0.4 to 1 and show the fraction of correct answers within each bin. The shaded band indicates the uncertainty based on the number of correct answers versus the total questions in each bin for each model. On the left panel, we show some of the strongest models in terms of calibration, including Claude-3.5-Sonnet, GPT-4o, Step-1, and Gemini-3.0-Opus. Claude-3.5-Sonnet demonstrates the best calibration, and models with higher accuracy often exhibit better calibration of their confidence, although they often show less confidence than they should. On the right panel, we display some of the weaker models in terms of calibration, including Gemini-3.0-Haiku, GPT-4, Gemini-1.5-Pro, and LLaMA-3-8B. Additional information on the performance on calibration for some other models are shown in Table~\ref{table7}.}
\label{fig10}
\end{figure*}

\begin{table*}[htbp]
\centering
\caption{Calibration Metrics of Various Large Language Models on Astronomy Multiple Choice Questions. The Pearson Correlation coefficient measures the linear correlation between model confidence and actual performance, with higher values indicating better relative calibration. The Mean Absolute Offset represents the average deviation of the model's confidence from its actual accuracy, with lower absolute values indicating more accurate absolute zero-point calibration. Bold values indicate the best performance in each metric. Claude-3.5-Sonnet shows the highest correlation, while Step-1 demonstrates the lowest mean absolute offset.}
\label{table7}
\begin{tabular}{lcc}
\hline
\textbf{Model} & \textbf{Pearson Correlation} & \textbf{Mean Absolute Offset} \\
\hline
Claude-3.5-Sonnet & \textbf{0.982} & 0.093 \\
Step-2 & 0.933 & 0.069 \\
Gemini-3.0-Opus & 0.928 & 0.077 \\
GLM-4-Air & 0.925 & 0.070 \\
Qwen-2-72B & 0.920 & 0.102 \\
LLaMA-3-70B & 0.918 & 0.090 \\
Gemma-2-27B & 0.917 & 0.041 \\
GPT-4o & 0.912 & 0.094 \\
Step-1 & 0.882 & \textbf{0.038} \\
Yi-Large & 0.872 & 0.067 \\
Gemini-3.0-Haiku & 0.806 & 0.091 \\
GLM-4-0520 & 0.801 & 0.060 \\
GPT-4 & 0.687 & 0.076 \\
Deepseek-v2 & 0.585 & 0.083 \\
Gemma-2-9B & 0.477 & 0.093 \\
Gemini-3.0-Sonnet & 0.424 & 0.107 \\
Qwen-2-7B & 0.281 & 0.133 \\
Gemini-1.5-Pro & 0.061 & 0.094 \\
LLaMA-3-8B & -0.463 & 0.123 \\
\hline
\end{tabular}
\end{table*}

\begin{figure*}
\centering
\includegraphics[width=1\textwidth]{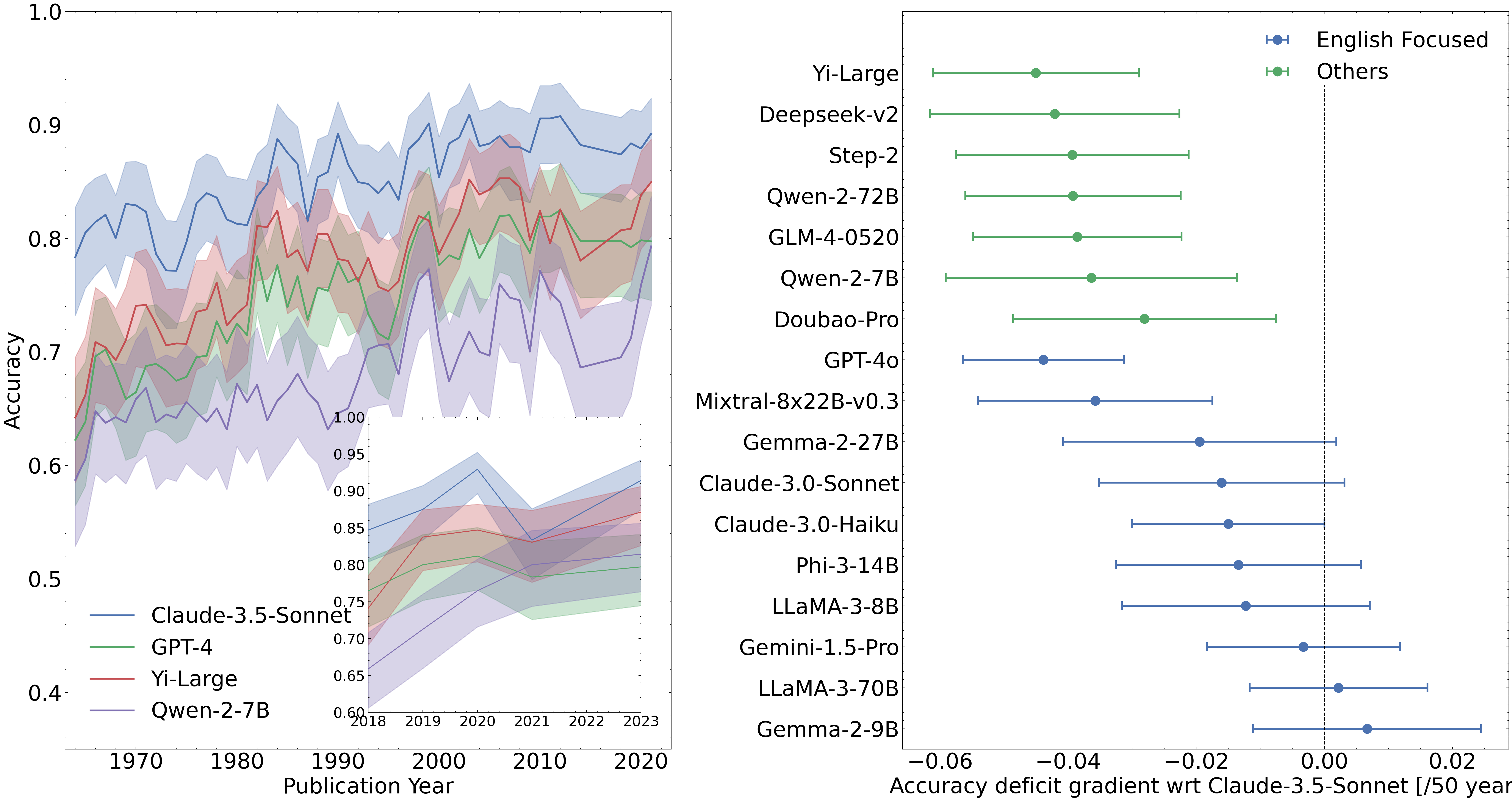}
\caption{Accuracy of Astronomy MCQ Answering over The Year of the Question. The left panel shows the accuracy of various LLMs in answering MCQs drawn from the Annual Review of Astronomy and Astrophysics, plotted as a function of publication year. We chose GPT-4 instead of GPT-4-turbo to demonstrate that the answering performance doesn't appear to be affected by significant question leakage. Despite GPT-4's training data cutoff around late 2021/early 2022, the model remains robust for questions from Annual Review articles published in 2022-2023. The inset provides a zoomed-in view of recent years. All models show improved performance on more recent questions. However, their ability to answer older questions from up to half a century ago varies by model. The right panel illustrates the difference in performance gradients between various models and Claude-3.5-Sonnet for questions from different years. A more negative value indicates a more pronounced degradation in accuracy for questions based on older Annual Reviews. English-foscused models (in blue) show a smaller decrease in performance for older questions, while non-English-focused models (in green), exhibit steeper negative gradients. But some English-focused models including Mixtral-8x22B-v0.1 and surprisingly GPT-4 also show a steeper degradation toward older questions.}
\label{fig11}
\end{figure*}

\subsection{Model Self-Awareness: Are You Sure?}

Given that even the best-performing models achieve only about 85\% accuracy (Claude-3.5-Sonnet correctly answering 5 out of 6 questions), it's crucial to assess whether models can gauge their own uncertainty. This implicit uncertainty calibration is critical for scientific research, as hallucination or overconfidence is unacceptable in the scientific community.

We tested the models' ability to gauge their confidence by prompting them to output probabilities for individual answers. And the confidence is defined as the maximum probability over the four possible answers divided by the sum of the four probabilities. Fig.~\ref{fig10} shows the number of correct answers as a function of the model's confidence for various proprietary models. We binned the confidence with a bin size of 0.1 from 0.4 to 1 and calculated the fraction of correct answers within each bin. The shaded band in the figure indicates the Wilson Score Interval, which we calculated based on the number of correct answers versus the total questions in each bin for each model. Ideally, perfect calibration would follow a 1:1 line, where the fraction of correct answers in any bin matches the model's confidence.

We note that although we could gauge the confidence through the logit of the output of the first token—i.e., the probability of A, B, C, and D, according to the next token prediction after the question. However, as we have discussed, we chose not to do that for two reasons: (a) weaker models tend to not be able to follow the instruction exactly, using the logit will bias against them; (b) more importantly, in most agent deployment scenarios, next token logit is often meaningless in real-life application. Being able to ask, through prompting, how confident the model is as a question matches real-life application more closely.

However, not all models perform equally well in this aspect. Some models, particularly smaller ones or earlier versions, show poorer calibration. We consider two metrics of calibration as shown in the two columns of Table~\ref{table7}: (1) We perform a weighted linear regression on the fraction of correct answers versus confidence. This `correlation' indicates whether a model has good self-awareness, demonstrating that it knows, in a relative sense, how accurate its answers are. (2) We also report the offset (intercept) of the linear regression above. An ideal case with a zero intercept would indicate perfect calibration of absolute confidence. We study the calibration performance for a subset of reprensentative models, and the results are shown in Table~\ref{table7}.

Even for slightly dated proprietary models, for instance, Gemini-1.5-Pro and Gemini-3.0-Sonnet have relatively low correlation coefficients (0.06 and 0.42 respectively), indicating less reliable self-assessment of their performance. This also demonstrates that it is only until most recently that calibration has generally improved. For example, we observe an improvement from GPT-4's correlation of 0.69 to GPT-4o's 0.91, showing significant progress in a short time span. The Claude-3.5-Sonnet even achieves a correlation score of 0.98, and 56 percentage point gain from Claude-3.0-Sonnet.

A particularly interesting trend for open-weights models is that while the $\gtrsim$30B models show reliable calibration, the calibration is much weaker for some of the smaller models. For example, LLaMA-3-70B achieves a strong Pearson correlation of 0.92, but this drops dramatically to -0.46 for LLaMA-3-8B (see also the right panel of Fig.~\ref{fig10}. And Qwen-2-7B drops the correlation to 0.28 from Qwen-2-72B's 0.92. Gemma models show the best compromise, with Gemma-2-27B achieving a correlation of 0.92 and Gemma-2-9B reaching 0.48.

Interestingly, we also observe a general trend across most models towards slight underconfidence—that is, the LLMs typically got more questions correct than they gave themselves credit for. This is evident in the mean absolute offset values in Table~\ref{table7}, where most models show positive offsets. For instance, Claude-3.5-Sonnet, despite its high accuracy, shows a mean absolute offset of 0.10, indicating a tendency to underestimate its performance. This trend could be interpreted in two ways. On one hand, it might suggest that the models are overly aligned to be cautious in their self-assessment, which isn't ideal for scientific applications where accurate self-evaluation is crucial. On the other hand, this underconfidence might point to limitations in our current benchmarking dataset. It's possible that a small fraction of the questions we generated are not formulated accurately or have somewhat ambiguous answers, leading even highly capable models to express uncertainty (see more question examples in Appendix~\ref{appendixC} and \ref{appendixD}). Notably, some models like Step-1 and Gemma-2-27B show better calibration with a lower mean absolute offset (0.04), suggesting that achieving better calibration is possible. These observations highlight the need for further refinement of our benchmarking approach. In our next paper, we aim to address these potential limitations by incorporating more detailed human curation of the questions and implementing methods for further self-cleaning of the data. This will help ensure that our benchmark provides an even more accurate assessment of LLM capabilities in astronomical research contexts.

Nevertheless, the strong performance in self-awareness is a promising feature for deploying these models in scientific research contexts, particularly for the top-performing models. However, the variation in calibration quality across different models underscores the importance of thoroughly evaluating this aspect when considering LLMs for astronomical research applications. The rapid improvement in calibration also suggests that future models may become even more reliable in assessing their own uncertainty, which is crucial for their application in scientific research.

\vspace{2cm}

\subsection{Assessing Potential Data Leakage and Temporal Performance}
\label{sec:temporal}

A critical concern in any benchmarking exercise is the potential for data leakage, where models may perform artificially well due to exposure to test data during training. This has been observed in previous benchmarks \citep{Huang2023}, leading to the development of new datasets like MMLU-pro \citep{Wang2024} and the use of exam questions released after model training to ensure fair evaluation.

Evaluating data leakage in our astronomical benchmark presents unique challenges because our aim to test models on factual knowledge and the historical evolution of astronomical literature inherently requires the use of historical data. Nonetheless, our questions are generated from the Annual Review of Astronomy and Astrophysics, not directly copied in verbatim from existing literature, which should have, to certain extent, mitigate this issue. To address these challenges and assess potential data leakage, we analyzed model performance as a function of the publication year of the source Annual Review articles. If models were simply recalling memorized knowledge from training data, we would expect to see a significant drop in performance for questions based on articles published after the model's training data cutoff date.

The left panel of Fig.~\ref{fig11} presents the accuracy of four representative models (Claude-3.5-Sonnet, GPT-4, Yi-Large, and Qwen-2-7B) as a function of the year in which the Annual Review associated with each question was published. The inset focuses on the years 2015-2024. We observe that there is no sudden drop in performance for recent years, suggesting models are not merely recalling training data. In fact all models perform better on questions from recent years compared to those from older years (e.g., 1960-1980), indicating a general understanding of concepts rather than simple recall. Also of particular interest is that, in the left panel, we specifically included the older version of GPT-4, instead of GPT-4o (with a training cutoff around late 2021/early 2022) to demonstrate its ability to perform well on questions based on post-cutoff articles.

The weaker performance on older reviews might not be surprising for two reasons. First, questions that are more dated and particularly related to certain space missions might be challenging even for human experts. Second, some fields might have subsequently led to drastically different points of view in certain research domains, and the questions extracted from those years might have led to answers that are not entirely accurate. This is also part of the reason why we decided to defer the full release of all the questions, though we emphasize that most questions remain robust.

Interestingly, agreeing with our evaluation in Fig.~\ref{fig4}, \ref{fig5}, \ref{fig8}, and \ref{fig9}, some models, depending on the training data adopted, can lead to weaker performance on questions that are more related to historical context. For example, on the left-hand panel in Fig.~\ref{fig11}, Yi-Large and GPT-4 (the older version) appear to show steeper gradients than Claude-3.5-Sonnet. This is further demonstrated in the right panel, where we fit the results from the left panel with linear regression models, taking into account the uncertainty, and calculate the difference between the slopes of the models in question versus the slope of Claude-3.5-Sonnet. The zero point here means that the models have a weak degradation for the older questions as demonstrated by Claude-3.5-Sonnet, and a more negative value means that the degradation is more prominent.

As shown, models that are not primarily focusing on English-language content, appear to have a weaker performance on older questions. The exact reason for that is hard to trace, especially for proprietary models, but this is perhaps not entirely surprising, since much of the training data from those developers can come from vast information that is not English-based literature, and might lead to slightly unfavorable performance for questions that are too tied to niche historical context. Interestingly, such stronger degradation does not only appear for the non-English-focused models but also for models like Mixtral-8x22B-v0.1 and the GPT-4 series. We hope this work sheds light on the potential avenues for improvements for various models in handling temporal aspects of astronomical knowledge.

\subsection{Balancing Performance and Affordability in LLM Deployment}

A key objective of our comprehensive benchmarking is to gauge the performance-efficiency trade-off, enabling efficient and affordable deployment of LLM agents for large-scale tasks previously deemed unfeasible. While the affordability threshold varies significantly by task, we can consider a concrete example to illustrate the challenges.

In a companion study (Sun et al., in prep.), we demonstrated the use of LLM agents working collaboratively to understand the spectral energy distribution of galaxies. This study showed that, on average, the entire reasoning process requires about 0.1 million tokens, with GPT-4o-level capabilities being necessary for robust reasoning and instruction following. For GPT-4o, this translates to approximately 1 USD per astronomical source. Considering cutting-edge space surveys like Euclid \citep{Laureijs2011} and Roman \citep{Wang2022,Troxel2023}, which aim to observe on the order of a billion sources, the inference cost could reach 1B USD (assuming no rate limitations). This cost is comparable to the construction cost of these telescopes. Realistically, an inference cost of 1-10\% of the build cost (1-2 orders of magnitude lower) would be more feasible for such projects.

The landscape for open-weights models presents its own challenges. Assuming GPT-4o or above capability is necessary (which is task-dependent), the current 70B parameter models and larger can achieve comparable performance - a remarkable feat in itself. However, the throughput for these models is relatively slow. Our testing shows a throughput of about 25 tokens per second per A100 GPU. Consequently, processing a billion sources at 0.1 million tokens each would require about 10 million GPU years, or a year of compute on a 10,000 GPU cluster - a scale currently unattainable in most academic settings. 

Looking ahead, three promising possibilities emerge that could address these challenges:

1. open-weights model improvements: open-weights models have shown potential even at the 7B parameter level. It remains plausible that careful fine-tuning strategies - whether through continual pretraining, specialized fine-tuning, or some form of direct preference optimization - could enable $<30$B models to achieve performance comparable to their $>30$B counterparts, with an order of magnitude better throughput. These have been the motivation to work on AstroLLaMA \citep{Nguyen2023} and AstroLLaMA-Chat \citep{Perkowski2024}. However, a significant challenge persists: smaller models tend to be more brittle, and improvements in astronomical recall might come at the cost of performance in other areas, even including the released AstroLLaMA and AstroLLaMA-Chat. We will explore this trade-off extensively in the coming papers.

2. The advancements of specialized computing units for Transformer models and API-based open-weights models: The rapid progress in specialized computing units and startups aiming to improve inference speed and cost of Transformer-based models has made it increasingly affordable to run APIs on open-weights models. In Fig.~\ref{fig2}, we highlight models that can routinely be found in such startups, including Qwen-2-7B, GLM-4-9B, Yi-1.5-9B, LLaMA-3-8B, LLaMA-3-70B, Qwen-2-72B, Yi-1.5-34B, InternLM-2.5-7B, Gemma-2-9B and Gemma-2-27B. We adopt the cost of Siliconflow\footnote{https://siliconflow.cn/pricing} as of late June 2024 for these models. As shown, Qwen-2-72B and LLaMA-3-70B are competitive models performing at the level of GPT-4o for this task, and can be run at an order of magnitude lower cost.

3. Rapid improvement in proprietary models: Our benchmark reveals a dramatic and universal improvement in price-performance ratios for proprietary models. Consistently across competitive proprietary models, we observe approximately a 10-fold improvement in pricing every 3-12 months for a given level of performance. If this trend continues, it projects an optimistic future for LLM deployment in astronomy. For simple LLM agent tasks, such applications could become entirely feasible by the end of this year, potentially at a cost of about $\mathcal{O}(1)\%$ of the build cost even for some of the most ambitious observational programs.

\section{Conclusion}

In this study, we present the first comprehensive benchmarking of proprietary and open-weights models released in the last 18 months since ChatGPT, focusing on their ability to handle doctoral-level astronomy questions. Our benchmarking dataset consists of 4,425 multiple choice questions derived from high-quality review articles in the Annual Review of Astronomy and Astrophysics. This evaluation is critical for understanding the robustness of current LLMs for deployment as research agents in astronomy.

Our main conclusions regarding proprietary models (including GPT, Claude, Gemini, GLM, DeepSeek, Step, Yi, Moonshot, ERNIE, ABAB, and Doubao) are as follows:

\begin{enumerate}
\item While some cutting-edge proprietary models have shown comparable reasoning abilities in other benchmarks, their performance can vary drastically in specialized domains like astronomical research.

\item Claude-3.5-Sonnet demonstrates the highest performance in our benchmark, achieving 85.0\% accuracy and outperforming GPT-4o (80.4\%) by 4.6 percentage points. Notably, Gemini-1.5-Pro (77.6\%) lags behind Claude-3.5-Sonnet by 7.4 percentage points in accuracy. Other top performers include Claude-3.0-Opus (82.7\%) and GPT-4o (80.4\%).

\item We observe a general trade-off of 3.5 percentage points in accuracy for every 10-fold increase in price within most given series of models. This substantial discrepancy between high-end models can result in three orders of magnitude difference in performance-to-cost ratio. This demonstrates that general benchmarking results do not necessarily translate into astronomy-specific performance, highlighting the importance of such specialized benchmarking studies.

\item Proprietary models show a strong dependency on the training data. For instance, non-English-focused models like GLM-4-0520 (75.1\%), Yi-Large (77.3\%), Doubao-Pro (70.1\%) and Deepseek-v2 (73.6\%), while proficient in other benchmarks, underperform in this scientific benchmark, particularly in areas like Historical and Theoretical Knowledge (e.g., GLM-4-0520 scores 75.0\% compared to Claude-3.5-Sonnet's 84.3\%) and Current Research and Advanced Topics, as well as topics such as Solar and Stellar Astrophysics, Earth and Planetary Astrophysics, and Instrumentation and Methods.

\item Most competitive proprietary models are currently improving at a rate equivalent to a 10-fold reduction in price for a given level of performance every 3-12 months.
\end{enumerate}

In terms of open-weights models, we have performed extensive benchmarking on the Phi, Mistral/Mixtral, Yi, LLaMA, Qwen, Gemma, GLM, InternLM and Deepseek series, and our conclusions are as follows:

\begin{enumerate}
\item open-weights models are catching up existing proprietary models in terms of detailed knowledge recall in astronomy, aligning with findings from other benchmarking efforts.

\item Models released before 2024, both at $<30$B and $>30$B levels, are significantly outperformed by proprietary models, with accuracy gaps as large as 20-30\%. The $<30$B models typically achieve 50-65\% accuracy, while the $>30$B models reach around 70\% accuracy.

\item However, models that appeared in the last few months since 2024 can reach the level of some proprietary models. The $>30$B or larger parameter scale models from LLaMA-3 (80.6\%), Qwen-2 (77.7\%), and Mixtral 8x22B (77.7\%) perform on par with or even surpass some proprietary models like Gemini-1.5-Pro (77.6\%). The $10-30$B level models (Phi-3-14B, Gemma-2-27B) now achieve 75.6\% accuracy.

\item Nonetheless, the latest proprietary models maintain a comfortable lead, with Claude-3.5-Sonnet still holding a 4.4 percentage point lead over the best open-weights model of LLaMA-3-70B.

\item We observe similar regional discrepancies in open-weights models, particularly in areas like Historical and Theoretical Knowledge and Current Research and Advanced Topics.
\end{enumerate}

Both latest proprietary and open-weights models exhibit excellent calibration and alignment, with top-performing models demonstrating high correlations ($>$0.9) between their confidence and actual performance. However, this is primarily true for the latest models; even slightly older versions show poorer calibration. Despite their decent performance, smaller open-weights models suffer severely from weaker calibration in confidence. We also note a general trend towards slight underconfidence for all models regardless of their sizes, with mean absolute offsets ranging from 0.04 (Step-1) to 0.13 (Qwen-2-7B), which may indicate either over-cautious alignment or potential limitations in our current benchmarking dataset.

Based on our benchmark, we suggest the following models to use at different price points are:

\begin{itemize}
\item For about U\$1 per 0.1M tokens: Claude-3.5-Sonnet (85.0\% accuracy, 98.2\% calibration)
\item For about U\$0.1 per 0.1M tokens: LLaMA-3-70B (80.6\% accuracy, 91.8\% calibration)
\item For about U\$0.01 per 0.1M tokens: Gemma-2-27B (75.6\% accuracy, 91.7\% calibration)
\item For about U\$0.001 or less per 0.1M tokens: Gemma-2-9B (71.5\% accuracy, 47.7\% calibration)
\end{itemize}

The combination of excellent error calibration, exponential price decreases in proprietary models, rising performance of open-weights LLMs, and the development of specialized computing units for fast, low-cost inference of open-weights models presents an additional pathway for affordable deployment of these models in astronomical research. It suggests that the current computational and financial barriers may be overcome sooner than initially anticipated, potentially revolutionizing our approach to large-scale data analysis in astronomy. As we continue to refine and deploy these models, we may be entering a new era of astronomy, where AI assistants significantly augment human researchers' capabilities, pushing the boundaries of our understanding of the universe.

\vspace{1cm}
This research was conducted using resources and services provided by the National Computational Infrastructure (NCI), which receives support from the Australian Government, and the Oak Ridge Leadership Computing Facility Frontier Nodes. We are also grateful for support from Microsoft's Accelerating Foundation Models Research (AFMR) program, which played a crucial role in enabling this benchmarking work. The work at Argonne National Laboratory was supported by the U.S. Department of Energy, Office of High Energy Physics and Advanced Scientific Computing Research, through the SciDAC-RAPIDS2 institute. Argonne National Laboratory is a U.S. Department of Energy Office of Science Laboratory operated by UChicago Argonne LLC under contract no. DE-AC02-06CH11357. The views expressed herein do not necessarily represent the views of the U.S. Department of Energy or the United States Government.

\bibliography{manuscriptNotes.bib}

\appendix 

\section{Inference Methodology Details}
\label{appendixA}

In our study, we used the following format for prompting the models to answer the multiple-choice questions:

\begin{questionbox}[grey]
\textbf{Prompt:}

You are an expert in general astrophysics. Your task is to answer and explain the following multiple-choice question on astrophysics, sourced from a dataset. The question is: \\

\textbf{Question}: [Question text] \\

Options: \\
\textbf{A}: [Option A] \\
\textbf{B}: [Option B] \\
\textbf{C}: [Option C] \\
\textbf{D}: [Option D] \\

Determine the correct answer using your astrophysics knowledge and provide a detailed explanation for why this answer is correct. \\

Ensure your explanation is thorough, clearly articulating your thought process based on astrophysical principles. \\

\textbf{Output format}: \\

\{ \\
    ``ANSWER": ``[The choice you decide to choose]", \\
    ``EXPLANATION": ``[Provide a valid explanation for the answer mentioned in ANSWER]" \\
\} \\

Give only one answer, either A, B, C or D, but not more than one, and always give an answer. \\

Adhere to the output format.
\end{questionbox}

When the system allowed for a separate system prompt, we used the first sentence (``You are an expert in general astrophysics.") as the system prompt.

Our experiments showed that the choice of sampling parameters (e.g., temperature) played a minor role in the performance of the models, especially for more recent and robust models. The key elements that consistently improved performance across models were: (1) The role-play framing the task as coming from an expert in astrophysics. (2) Requesting both an answer and an explanation, encouraging a chain of thought inference. (3) Providing a clear output format. Although the third point was not often adhered to, which required us to further parse the results with GPT-4o. This observation aligns with the broader trend in LLM research, where more advanced models show increasing robustness to prompt variations, as long as the essential task information is conveyed.
\begin{figure*}
\centering
\includegraphics[width=1.0\textwidth]{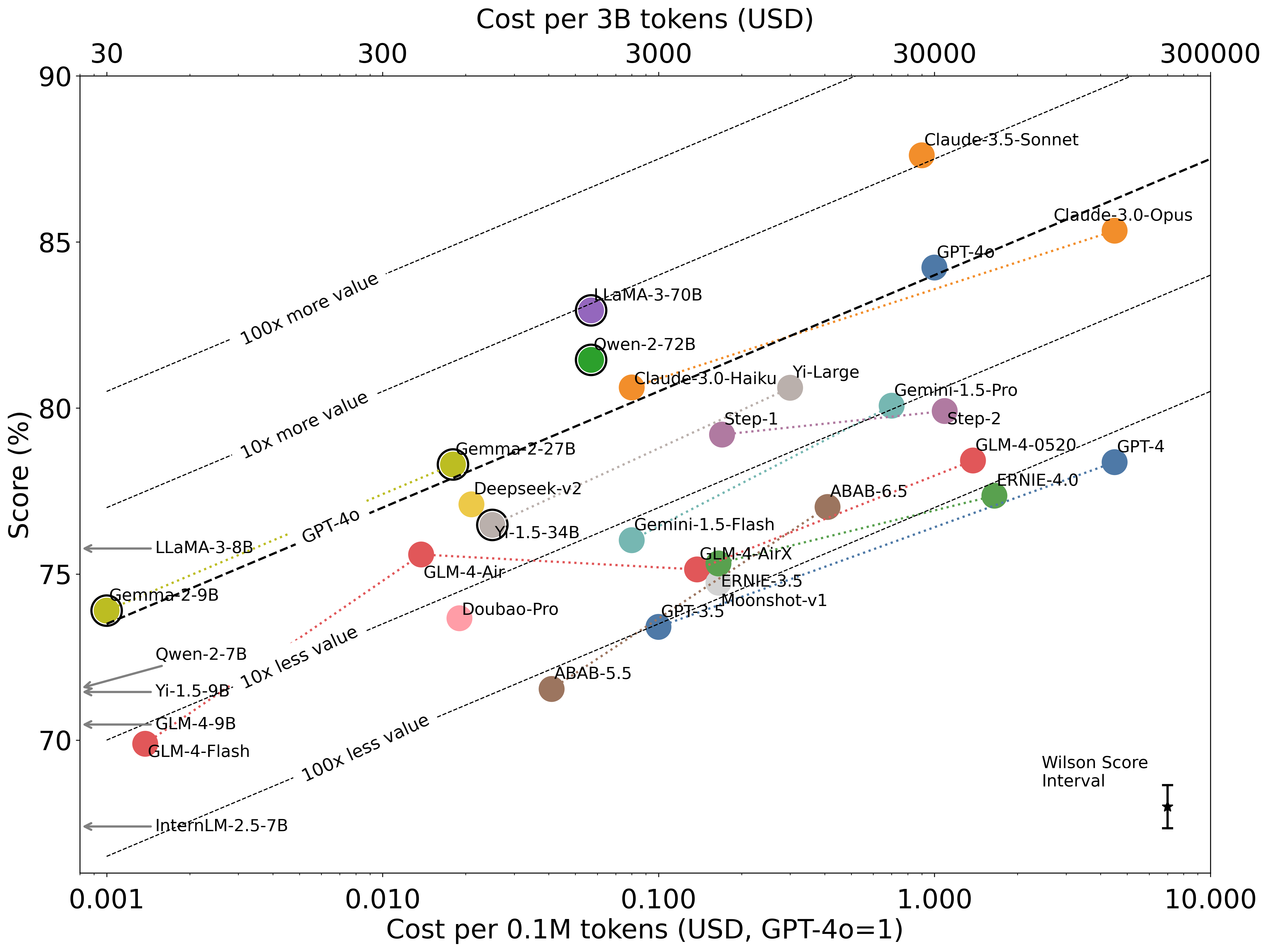}
\caption{Performance on Recent Astronomy MCQs (Post-1990) vs. Cost. This plot resembles Fig.~\ref{fig2} but focuses solely on MCQ benchmark questions from Annual Review articles published after 1990. Some LLMs struggle with questions involving historical context, potentially affecting their scores. By limiting the analysis to post-1990 questions, we see the performance gap between models narrow. However, the relative ranking of models remains largely unchanged. This suggests that while questions from older literature partly explain the differences between models, the rankings shown in Fig.~\ref{fig2} are robust. For open-weights models (marked with black circles), we use pricing from the SiliconFlow API as of June/July 2024. The dashed lines represent the price-efficiency trade-off (see text for details), with the bold dashed line showing GPT-4o's value.}
\label{fig12}
\end{figure*}

\section{Performance Analysis on Post-1990 Questions}
\label{appendixB}

As we have demonstrated in the right panel Fig.~\ref{fig11} (also see Fig.~\ref{fig4}, Fig.~\ref{fig5}, Fig.~\ref{fig8}, Fig.~\ref{fig9}), some models perform worse than their competitors due to differences in training data, particularly in questions related to historical context. These differences might have less relevance to the models' ability to execute tasks as LLM agents in astronomical research. Additionally, some questions extracted from Annual Reviews articles dating back almost half a century ago may contain outdated information or reflect outdated consensus, potentially affecting the validity of the answers.

To mitigate these biases, Fig.~\ref{fig12} examines the case where we restrict our analysis to questions extracted from Annual Reviews published after 1990. This leaves us with 2,245 questions out of the original 4,425 questions. As shown in Fig.~\ref{fig12}, most models perform better when evaluated on this subset of more recent questions. For example: Claude-3.5-Sonnet improves from 85.0\% to 87.6\% (+2.6 points), GPT-4o improves from 80.4\% to 84.2\% (+3.8 points), LLaMA-3-70B improves from 80.6\% to 83.0\% (+2.4 points) and Qwen-2-72B improves from 77.7\% to 81.5\% (+3.8 points)

The performance gap between models narrows slightly. For instance, the difference between Claude-3.5-Sonnet and GPT-4o decreases from 4.6 percentage points to 3.4 percentage points. This suggests that some of the performance differences observed in the full dataset may indeed be partly attributed to varying abilities in handling historical context or older scientific consensus.

However, despite these changes, the relative performance of models as shown in Fig.~\ref{fig12} remains largely consistent with that shown in Fig.~\ref{fig2}. This consistency suggests that while historical context-related data might have impacted some models' performance, the overall evaluation of model strengths in this study remains largely robust and reliable. The observed improvements across all models when focusing on more recent questions highlight the importance of considering the temporal aspect of scientific knowledge when evaluating LLMs for specialized domains like astrophysics.

\section{Classification of Astrophysics Questions}
\label{appendixC}

In this study, as shown in Table~\ref{table3} and \ref{table4} we categorized the questions according to the six main subclasses used in the arXiv's astro-ph section. This classification system reflects the broad areas of research within astrophysics: (1) Solar and Stellar Astrophysics, (2) Earth and Planetary Astrophysics, (3) Astrophysics of Galaxies, (4) Cosmology and Nongalactic Astrophysics, (5) High Energy Astrophysics, and (6) Instrumentation and Methods for Astrophysics. This categorization allows for a more nuanced analysis of model performance across different specializations. We'll provide two examples for each class to showcase the types of questions that appear in each. The full benchmarking dataset will be released in our upcoming paper.

\subsection{Solar and Stellar Astrophysics}

This category focuses on the physics of stars, including our Sun, their formation, evolution, and internal processes.

\begin{questionbox}[boxcolor1]
\textbf{Paper ID:} 1971ARA\&A...9..127S \\

\textbf{Question:} What is the most likely range of initial masses for stars that become planetary nebulae? \\

(A) 0.5 to 1.0 solar masses \\
(B) 1.0 to 1.5 solar masses \\
(C) 1.5 to 2.0 solar masses \\
(D) 2.0 to 2.5 solar masses \\

\textbf{Correct Answer:} B \\

\textbf{Explanation:} The article discusses the spatial and velocity distribution of planetary nebulae, suggesting they originate from stars with masses similar to those in the disk population. This population typically has stars with masses between 1.0 and 1.5 solar masses. Additionally, the article mentions that the rate of planetary nebula formation aligns with the rate of white dwarf creation, further supporting this mass range as white dwarfs usually originate from stars within this mass range.
\end{questionbox}

\begin{questionbox}[boxcolor1]
\textbf{Paper ID:} 2003ARA\&A..41..465H \\

\textbf{Question:} How do the masses of cool white dwarfs compare to those of hot white dwarfs, and what are the potential implications? \\

(A) Cool white dwarfs generally have lower masses, suggesting that they originate from lower-mass progenitor stars. \\
(B) Cool white dwarfs have a similar mass distribution to hot white dwarfs, indicating a consistent mass range for white dwarf progenitors. \\
(C) Cool white dwarfs tend to have higher masses, possibly due to their older ages and origin from higher-mass progenitors. \\
(D) The mass distribution of cool white dwarfs is bimodal, with distinct populations of low-mass and high-mass white dwarfs. \\

\textbf{Correct Answer:} C \\

\textbf{Explanation:} The article presents evidence that cool white dwarfs exhibit a slightly higher mean mass compared to hot white dwarfs. This observation is consistent with the idea that older white dwarfs, which are cooler, likely originate from higher-mass progenitor stars that have had more time to evolve and reach the white dwarf stage.
\end{questionbox}

\subsection{Earth and Planetary Astrophysics}
This category encompasses the study of planets, both in our solar system and exoplanets, as well as planetary formation and evolution.

\begin{questionbox}[boxcolor2]
\textbf{Paper ID:} 2019ARA\&A..57..617M \\

\textbf{Question:} What are the primary advantages of directly imaged spectra of exoplanets compared to transmission spectra? \\

(A) Directly imaged spectra typically have higher resolution and signal-to-noise ratio, allowing for more precise measurements of atmospheric features. \\
(B) Directly imaged spectra provide information about the planet's atmospheric composition at all orbital phases, while transmission spectra only probe the day-night terminator region. \\
(C) Direct imaging allows for the detection of exoplanets at much smaller orbital separations than transit spectroscopy. \\
(D) Directly imaged spectra are less affected by the presence of clouds and hazes in the planet's atmosphere, leading to more accurate abundance estimates. \\

\textbf{Correct Answer:} A \\

\textbf{Explanation:} The article explains in section 2.3 that directly imaged spectra often have higher resolution and signal-to-noise ratio due to the use of large-aperture ground-based telescopes with adaptive optics. This allows for more precise measurements of atmospheric features. However, directly imaged spectra are typically obtained at a single, unknown orbital phase and are more affected by degeneracies due to the unknown mass, radius, and gravity of the planet.
\end{questionbox}

\begin{questionbox}[boxcolor2]
\textbf{Paper ID:} 1993ARA\&A..31..523M \\

\textbf{Question:} How does the quasi-geostrophic (QG) model explain the presence of multiple vortices at the same latitude on Jupiter, such as the chain of cyclones and anti-cyclones observed at 41$^\circ$S? \\

(A) Multiple vortices at the same latitude are inherently unstable and will eventually merge in the QG model. \\
(B) The QG model suggests that the presence of multiple vortices is due to the influence of Jupiter's moons. \\
(C) The QG model predicts the formation of K\'arm\'an vortex streets, where staggered rows of cyclones and anti-cyclones prevent merging due to their mutually repulsive interactions. \\
(D) The multiple vortices are remnants of a larger vortex that broke apart due to turbulence, and they will eventually re-merge. \\

\textbf{Correct Answer:} C \\

\textbf{Explanation:} The article proposes that the observed co-existence of multiple vortices at the same latitude can be explained by the formation of K\'arm\'an vortex streets. These structures consist of staggered rows of cyclones and anti-cyclones, where the cyclones act as `blocking vortices', preventing the anti-cyclones from merging due to their mutually repulsive interactions. This arrangement is consistent with observations, such as the chain of vortices at 41$^\circ$S, where cyclones and anti-cyclones are observed in close proximity without merging.
\end{questionbox}

\subsection{Astrophysics of Galaxies}

This category deals with the structure, formation, and evolution of galaxies, including our Milky Way.

\begin{questionbox}[boxcolor3]
\textbf{Paper ID:} 1980ARA\&A..18..165M \\

\textbf{Question:} What is the most likely explanation for the observed alignment between the radio axis of double radio sources and the minor axis of their host elliptical galaxies? \\

(A) The alignment is purely coincidental and has no physical basis. \\
(B) The radio jets preferentially expand along the path of least resistance, which coincides with the minor axis of the galaxy due to its flattened shape. \\
(C) The radio axis is aligned with the angular momentum axis of the galaxy, suggesting a connection between the central engine and the galaxy's rotation. \\
(D) The alignment is caused by the interaction of the radio jets with the surrounding intergalactic medium, which shapes the radio source along the minor axis of the galaxy. \\

\textbf{Correct Answer:} C \\

\textbf{Explanation:} The article, particularly in section 5.3, discusses the observed alignment between radio and galaxy axes and argues that it suggests a link between the radio-source production axis and the angular momentum axis of the galaxy. This implies a connection between the central engine (likely a black hole) and the galaxy's rotation. Options B and D, while potentially contributing factors, do not address the fundamental relationship between the radio source and the galaxy's intrinsic properties.
\end{questionbox}

\begin{questionbox}[boxcolor3]
\textbf{Paper ID:} 1963ARA\&A...1..149R \\

\textbf{Question:} How can the study of integrated spectra of galaxies provide information about their stellar content? \\

(A) The overall spectral type of a galaxy, determined from its integrated spectrum, directly corresponds to the most common type of star in the galaxy. \\
(B) The presence and strength of specific absorption lines in the integrated spectrum can reveal the types of stars contributing most to the galaxy's light at different wavelengths. \\
(C) The redshift of the integrated spectrum indicates the galaxy's age, which in turn provides information about the types of stars present. \\
(D) The shape of the continuum in the integrated spectrum reflects the distribution of dust in the galaxy, which can be used to infer the types of stars that are obscured. \\

\textbf{Correct Answer:} B \\

\textbf{Explanation:} The article explains that the presence and strength of specific absorption lines in the integrated spectrum of a galaxy can be used to identify the types of stars that contribute most to the galaxy's light at different wavelengths. For example: Cyanogen bands indicate the presence of late G and early K giant stars, weak TiO bands suggest a lack of late-type giants or supergiants, and the strength of Mg b blend, Na D lines, and CaI line points to a significant population of late-type dwarf stars. By analyzing these features across a wide wavelength range, astronomers can construct a qualitative picture of the galaxy's Hertzsprung-Russell diagram and its spectrum-luminosity function.
\end{questionbox}

\subsection{Cosmology and Nongalactic Astrophysics}

This category covers the study of the universe as a whole, its origin, evolution, and large-scale structures, as well as active galactic nuclei (AGN) and other extragalactic phenomena.
\begin{questionbox}[boxcolor4]
\textbf{Paper ID:} 1998ARA\&A..36..267R \\

\textbf{Question:} How does the column density distribution function (CDDF) provide information about the Lyman alpha forest? \\

(A) The CDDF describes the number of absorbers per unit column density interval. Its power-law shape over a wide range of column densities suggests a common origin for most Lyman alpha absorbers and provides insights into the underlying physical processes. \\
(B) The CDDF measures the clustering of Lyman alpha absorbers along the line of sight, indicating the presence of large-scale structures in the intergalactic medium. \\
(C) The CDDF describes the redshift evolution of the Lyman alpha forest, showing how the number of absorbers changes with cosmic time. \\
(D) The CDDF measures the temperature of the intergalactic medium based on the Doppler broadening of absorption lines. \\

\textbf{Correct Answer:} A \\

\textbf{Explanation:} The CDDF describes the number of absorbers per unit column density interval. Its power-law shape over a wide range of column densities suggests a common origin for most Lyman alpha absorbers and provides information about the underlying physical processes. The article explains this in detail in section 3.4.
\end{questionbox}

\begin{questionbox}[boxcolor4]
\textbf{Paper ID:} 2003ARA\&A..41..645R \\

\textbf{Question:} What is the primary advantage of using weak gravitational lensing to map dark matter compared to methods that rely on the distribution of light? \\

(A) Weak lensing is less affected by dust obscuration, providing a clearer view of dark matter. \\
(B) Weak lensing directly measures the mass distribution, including dark matter, rather than just the distribution of luminous matter. \\
(C) Weak lensing is sensitive to the motion of dark matter, revealing its dynamic behavior. \\
(D) Weak lensing can probe dark matter at higher redshifts than traditional methods. \\

\textbf{Correct Answer:} B \\

\textbf{Explanation:} Unlike methods that rely on the distribution of light, which only traces luminous matter, weak lensing directly measures the total mass distribution, including dark matter. This is because the bending of light by gravitational lensing is caused by the total mass, regardless of its luminosity. This direct measurement allows for a more accurate and reliable mapping of dark matter compared to methods that require assumptions about the relationship between light and mass.
\end{questionbox}

\subsection{High Energy Astrophysics}

This category involves the study of high-energy phenomena in the universe, including black holes, neutron stars, and cosmic rays.

\begin{questionbox}[boxcolor5]
\textbf{Paper ID:} 2009ARA\&A..47..107H \\

\textbf{Question:} How do the spins of black holes in a binary system affect the gravitational waves they emit? \\

(A) The spins of black holes have a negligible impact on the emitted gravitational waves, as the dominant contribution comes from the mass quadrupole moment. \\
(B) Spin-orbit and spin-spin interactions introduce additional parameters into the waveform, encoding information about the binary's masses and spins, but these parameters may be highly correlated. \\
(C) Spin precession, caused by the motion of the black holes in the binary's curved spacetime, leads to phase and amplitude modulation of the gravitational waves, breaking parameter degeneracies and enhancing our ability to characterize the system. \\
(D) The presence of spin significantly reduces the recoil or `kick' experienced by the black hole due to the emission of gravitational waves. \\

\textbf{Correct Answer:} C \\

\textbf{Explanation:} The spins of black holes in a binary system have a profound effect on the emitted gravitational waves. Spin-orbit and spin-spin interactions introduce additional parameters into the waveform, but the most significant impact comes from spin precession. As the black holes' spins precess due to their motion in the binary's curved spacetime, the gravitational waves exhibit phase and amplitude modulation. This modulation breaks parameter degeneracies, allowing for more precise measurements of the binary's properties, such as the individual masses and spins of the black holes.
\end{questionbox}

\begin{questionbox}[boxcolor5]
\textbf{Paper ID:} 1993ARA\&A..31..175M \\

\textbf{Question:} What was the primary energy source for the luminosity of SN 1987A after the initial 3 months? \\

(A) Nuclear fusion in the core \\
(B) Gravitational contraction \\
(C) Radioactive decay of elements like Cobalt-56 \\
(D) Accretion onto a central compact object \\

\textbf{Correct Answer:} C \\

\textbf{Explanation:} The article clearly states that after the initial 3 months, during which the trapped radiation diffused out, the light curve of SN 1987A closely followed the decay rate of Cobalt-56, indicating radioactive decay as the primary energy source. This is further supported by the discussion on the role of other radioactive isotopes like Cobalt-57 and Titanium-44 in powering the later stages of the light curve.
\end{questionbox}

\subsection{Instrumentation and Methods for Astrophysics}

This category focuses on the development and application of tools and techniques used in astronomical observations and data analysis.

\begin{questionbox}[boxcolor6]

\textbf{Paper ID:} 1999ARA\&A..37...65R \\

\textbf{Question:} What was the main point of contention between radio astronomers and the operators of the GLONASS satellite navigation system? \\

(A) The GLONASS satellites were transmitting signals within a radio astronomy band allocated for observations of the hydroxyl (OH) molecule, causing significant interference. \\
(B) The GLONASS satellites were physically obstructing the view of radio telescopes, hindering astronomical observations. \\
(C) The GLONASS system was consuming an excessive amount of radio frequency spectrum, leaving insufficient bandwidth for radio astronomy research. \\
(D) The GLONASS satellites were generating harmful radiation that posed a health risk to astronomers working at radio observatories. \\

\textbf{Correct Answer:} A \\

\textbf{Explanation:} The article details the conflict between radio astronomers and the operators of the GLONASS system, which arose due to the GLONASS satellites transmitting signals within the 1610.6-1613.8 MHz band allocated for radio astronomy observations of the hydroxyl (OH) molecule. This interference significantly impacted radio astronomy research, leading to extensive negotiations and efforts to resolve the issue.
\end{questionbox}

\begin{questionbox}[boxcolor6]
\textbf{Paper ID:} 1982ARA\&A..20..367W \\

\textbf{Question:} What is the main advantage of using `speckle interferometry'? \\

(A) It allows for observations at longer wavelengths, such as the infrared. \\
(B) It is a simple and inexpensive technique compared to other methods. \\
(C) It can achieve higher angular resolution than traditional imaging techniques. \\
(D) It is less affected by atmospheric seeing compared to direct imaging. \\

\textbf{Correct Answer:} C \\

\textbf{Explanation:} Speckle interferometry is a technique that analyzes the short-exposure speckle patterns in a star image to extract high-resolution information about the object. The article explains that this method can overcome the limitations imposed by atmospheric seeing and achieve diffraction-limited resolution, making it particularly valuable for studying close binary stars and resolving the structures of distant galaxies.
\end{questionbox}

\section{Classification of Astrophysics Questions by Ability}
\label{appendixD}

In addition to categorizing questions by topic, we also classified them based on the cognitive abilities and skills they require. This classification allows for a more nuanced analysis of model performance across different aspects of astrophysical knowledge. The categories are as follows:

\subsection{Understanding Fundamental Concepts}

This category tests the grasp of basic principles and foundational knowledge in astrophysics.

\begin{questionbox}[boxcolor1]
\textbf{Paper ID:} 1966ARA\&A...4..145M \\

\textbf{Question:} What is the significance of the polarization of radio emission from radio galaxies? \\

(A) It indicates the presence of a strong, ordered magnetic field throughout the source \\
(B) It provides information about the distribution and orientation of the magnetic field in the source \\
(C) It is a direct measure of the energy of the relativistic electrons \\
(D) It is primarily caused by scattering of radio waves by dust grains \\

\textbf{Correct Answer:} B \\

\textbf{Explanation:} The article discusses how the polarization of radio emission, particularly the direction of polarization in different parts of the source, can be used to infer the structure of the magnetic field. The generally low degree of polarization observed suggests that the magnetic field is not uniform in direction. By studying the polarization, astronomers can gain insights into the magnetic field configuration within radio galaxies.
\end{questionbox}

\subsection{Technical and Observational Techniques}

This category focuses on specific methods, tools, and techniques used in astrophysical research and observations.

\begin{questionbox}[boxcolor2]
\textbf{Paper ID:} 1970ARA\&A...8..265H \\

\textbf{Question:} What is the primary factor that limits the sensitivity of radio telescopes when searching for pulsars at low frequencies? \\

(A) The intrinsic faintness of pulsars at low frequencies. \\
(B) Radio interference from terrestrial sources. \\
(C) The dispersion of pulsar signals by the interstellar medium, requiring smaller bandwidths and leading to a loss of sensitivity. \\
(D) The limited collecting area of existing radio telescopes. \\

\textbf{Correct Answer:} C \\

\textbf{Explanation:} As detailed in the `Search techniques and limitations' section, the dispersion of pulsar signals by the interstellar medium causes a frequency sweep. To observe this at low frequencies requires very small receiver bandwidths to avoid smearing out the signal, which in turn reduces sensitivity. This effect becomes more pronounced with increasing dispersion measure and hence distance.
\end{questionbox}

\subsection{Analytical and Reasoning Skills}

This category includes questions that require critical thinking, problem-solving, and the application of knowledge to analyze and interpret data or phenomena.

\begin{questionbox}[boxcolor3]
\textbf{Paper ID:} 1990ARA\&A..28..437H \\

\textbf{Question:} How do quasar-galaxy correlations inform us about the environments of quasars? \\

(A) Studies of quasar-galaxy correlations suggest that low-redshift quasars are more likely to have companion galaxies than typical galaxies, but they do not usually reside in rich clusters. The environments of high-redshift quasars are less clear, with some studies suggesting positive correlations and others indicating anticorrelations or no association. \\
(B) Quasar-galaxy correlations provide evidence for the non-cosmological nature of quasar redshifts, as quasars appear to be associated with galaxies at much lower redshifts. \\
(C) Quasar-galaxy correlations are used to study the phenomenon of gravitational lensing, where the light from a distant quasar is bent by the gravity of a foreground galaxy. \\
(D) Quasar-galaxy correlations are not a reliable tool for studying quasar environments due to the low space density of quasars and the challenges of identifying true associations. \\

\textbf{Correct Answer:} A \\

\textbf{Explanation:} Section 4.2 discusses quasar-galaxy correlations and their implications for quasar environments. The article highlights the findings from various studies, indicating that low-redshift quasars tend to have more companion galaxies than average but are not typically found in rich clusters. The situation at high redshifts is more complex and requires further investigation.
\end{questionbox}

\subsection{Historical and Theoretical Knowledge}

This category tests the understanding of historical developments, theoretical models, and key discoveries in the field of astrophysics.

\begin{questionbox}[boxcolor4]
\textbf{Paper ID:} 1994ARA\&A..32..319W \\

\textbf{Question:} What was the primary motivation for considering non-baryonic dark matter in the context of CMB anisotropies? \\

(A) The need to explain the observed large-scale structure in the universe. \\
(B) The inability of baryonic matter models to account for the observed CMB anisotropies. \\
(C) Claims of a significant mass for the electron neutrino, suggesting a potential dark matter candidate. \\
(D) The need to reconcile the age of the universe with the observed expansion rate. \\

\textbf{Correct Answer:} C \\

\textbf{Explanation:} In the early 1980s, claims of a substantial mass for the electron neutrino spurred interest in non-baryonic dark matter models. This was because such a massive neutrino could potentially account for the missing mass needed to explain the observed gravitational effects in the universe. As a result, researchers started exploring the implications of such models for CMB anisotropies.
\end{questionbox}

\subsection{Current Research and Advanced Topics}

This category focuses on recent research, advanced topics, and cutting-edge developments in astrophysics.

\begin{questionbox}[boxcolor6]
\textbf{Paper ID:} 2019ARA\&A..57..335F \\

\textbf{Question:} What is the role of gravitational screening mechanisms in modified gravity theories? \\

(A) Screening mechanisms suppress the effects of modified gravity on small scales, such as within galaxies and clusters, while allowing them to manifest on cosmological scales. \\
(B) Screening mechanisms enhance the strength of gravity in regions of high density, leading to the formation of black holes and other compact objects. \\
(C) Screening mechanisms explain the observed acceleration of the universe's expansion without the need for dark energy. \\
(D) Screening mechanisms account for the observed discrepancies between the predicted and measured values of the Hubble constant. \\

\textbf{Correct Answer:} A \\

\textbf{Explanation:} The article discusses how gravitational screening mechanisms can suppress the effects of modified gravity in regions of high density or curvature, such as within galaxies and clusters, while allowing them to manifest on larger, cosmological scales. This is achieved by suppressing the fifth force charge or coupling in these high-density environments. 
\end{questionbox}

These examples demonstrate the diverse range of abilities and skills required in astrophysics, from understanding fundamental concepts to analyzing complex phenomena and engaging with cutting-edge research. By categorizing questions in this manner, we can assess how well different models perform across various aspects of astrophysical knowledge and reasoning.

It's important to note that unlike the topic-based categorization, the classification by ability is arguably more subjective and potentially ambiguous. Many questions may test multiple abilities simultaneously, making a clear-cut categorization challenging. For this study, we assigned each question to a single ability category that we deemed most prominent, acknowledging that this approach is undoubtedly a simplification of the complex nature of these questions. This crude categorization allows us to discern broad trends in model performance across different cognitive skills. However, we recognize the limitations of this approach and aim to explore more nuanced categorization methods in future work. Our upcoming paper will investigate into this issue, presenting a more detailed human curation and analysis of these questions to provide a more comprehensive understanding of how different abilities are represented and tested in astrophysical knowledge assessment.

\bibliographystyle{aasjournal}

\end{CJK*}
\end{document}